\documentclass[preprint,notitlepage]{revtex4-1}

\usepackage{amsmath}
\usepackage[pdftex]{graphicx}
\usepackage{epstopdf}

\def \bn{\begin{align}}
\def \en{\end{align}}
\def \be{\begin{equation}}
\def \ee{\end{equation}}
\def \bea{\begin{eqnarray}}
\def \eea{\end{eqnarray}}
\def \ba{\begin{array}}
\def \ea{\end{array}}

\def \e{{\epsilon}}
\def \a{{\alpha}}

\def \d{{\delta}}
\def \w{{\omega}}

\def \D{{\Delta}}

\def \etal {{\it et al. }}

\def \nn{{\nonumber}}

\usepackage{ulem}

\def \appendix{{}}

\begin{document}
%\vspace{0.5cm}

\title{Exploring Quasiparticles in High-$\mathbf{T_c}$ Cuprates Through Photoemission, Tunneling, and X-ray Scattering Experiments}

\author{Emanuele G. Dalla Torre*, Yang He, David Benjamin, Eugene Demler}
\affiliation{Department of Physics, Harvard University, Cambridge, MA 02138\\}

\maketitle

{\bf One of the key challenges in the field of high-temperature superconductivity is understanding the nature of fermionic quasiparticles. Experiments consistently demonstrate the existence of a second energy scale, distinct from the d-wave superconducting gap, that persists above the transition temperature into the ``pseudogap'' phase. One common class of models relates this energy scale to the quasiparticle gap due to a competing order, such as the incommensurate ``checkerboard'' order observed in scanning tunneling microscopy (STM) and resonant elastic X-ray scattering (REXS). In this paper we show that these experiments are better described by identifying the second energy scale with the inverse lifetime of quasiparticles. We develop a minimal phenomenological model that allows us to quantitatively describe STM and REXS experiments and discuss their relation with photoemission spectroscopy. Our study refocuses questions about the nature of the pseudogap phase to the study of the origin of inelastic scattering.}
\newpage

Understanding the elementary excitations of superconducting cuprates is one of the central problems in the field of high-$T_c$ superconductivity. It is widely accepted that the quasiparticle spectrum involves two distinct energy scales: the superconducting gap and the ``pseudogap'' (see for example Refs.~\cite{hufner08, shen11} and references therein). However, the physical origin of the pseudogap is still debated. One common interpretation relates the pseudogap to a distinct long-range order that competes with superconductivity. Supporting evidence for this order was provided by periodic modulations in scanning tunneling microscope (STM) maps  \cite{hoffman02B,kapitulnik03,davis04,yazdani04,mcelroy05,davis07} and pronounced peaks in X-ray scattering  \cite{abbamonte05,ghiringhelli12,chang12,damascelli13,yazdani13B}. The simplest interpretation of both experiments is the presence of an incommensurate charge-density-wave (CDW) order coexisting with superconductivity. This competition can be described in terms of first-principles two-gap theories (see for example Ref.~\cite{zhang02,sachdev03}). However, two gap models do not provide an adequate description of all experimental observations. Motivated by angle-resolved photoemission spectroscopy (ARPES) \footnote{See in particular Norman \etal \cite{kanigel07}, Reber \etal \cite{dessau12}, and \appendix \ref{sec:ARPES}} and electrical conductivity measurements \cite{hussey09}, in this paper we explore a scenario in which the second energy scale characterizing the pseudogap phase is the finite inelastic relaxation rate of antinodal quasiparitcles. 
%We show that Friedel oscillations of quasiparitcles with finite lifetime lead to incommensurate peaks in both STM and X-ray signals, of the same character as observed in experiments. 
We develop a simple phenomenological model that accurately describes the experimental observations and, in particular, accounts for the wavevector and correlation length of the spatial modulations.

Our interpretation of the experimental results relies on a detailed theoretical analysis of the interplay between finite quasiparticle lifetime and disorder in $d$-wave superconductors. Even in the absence of true CDW order, Friedel oscillations around a single impurity can give rise to short-range incommensurate checkerboard patterns. For materials with a long quasiparticle lifetime, these oscillations are well described by the ``octet model''  \cite{lee03} and appear as dispersive peaks in the STM spectra  \cite{hoffman02A}. In contrast, when the quasiparticle lifetime is short, we find that the STM spectra exhibit non-dispersive peaks close to the antinodal scattering wavevectors. The predicted signal for both STM and REXS agrees quantitatively with recent experiments on 
$\mathrm{(Pb_x,Bi_{2-x})_2(La_y,Sr_{2-y})CuO_{6+\delta}}$ (Pb-Bi2201)  \cite{hudson08,yanghe13,damascelli13}, 
$\mathrm{Bi_2Sr_2CaCu_2O_{8+\delta}}$ (Bi2212)  \cite{davis08_Mott,alldredge08,schmidt11,fujita11,yazdani13B},
and $\mathrm{YBa_2Cu_3O_{7-x}}$ (Y123)  \cite{ghiringhelli12}. 

The present analysis does not rule out the existence of competing orders in cuprates. Some materials (such as $\mathrm{La_{2-x}Sr_xCuO_2}$ at high magnetic field  \cite{lake02} and $\mathrm{La_{2-x}Ba_xCuO_4}$ at $1/8$ filling  \cite{abbamonte05}) display sharp diffraction peaks, accompanied by a suppression of the superconducting critical temperature $T_c$. These phenomena indicate the onset of a true long-range order and require a separate analysis  \cite{kivelson03,berg09}. Moreover, the enhanced inelastic scattering of quasiparticles generally observed in underdoped samples can be due to fluctuations of a competing order \cite{sachdev13}, associated for example with a point-group symmetry breaking \cite{kivelson13}. However we show that Friedel oscillations of quasiparticles with finite lifetime are consistent with all experimental findings, regardless of the microscopic origin of inelastic scattering. Thus the key question for future experiments is to understand the physical origin of strong inelastic scattering of quasiparticles.

The starting point of our analysis of fermionic quasiparticles in cuprates is the retarded Green's function $G(k,\w)$. In the absence of disorder, and using the Nambu notation (see Ref.~\cite{ambegaokar69} for an introduction), it satisfies
\be G^{-1}(k,\w) = \left(\ba{c c}\w - \e_k +\mu + i \Gamma_k & \D_k \\ \D_{-k} & \w + \e_{-k} -\mu + i \Gamma_{-k} \ea\right)\;.~\label{eq:G0}\ee
The four ingredients of Eq.~(\ref{eq:G0}) are: (i) a phenomenological band structure $\e_k$, obtained from ARPES measurements (we use the single-band model of Ref.~\cite{campuzano95} for both Bi2212 and Pb-Bi2201, and the two-band model of Ref.~\cite{shen98} for Y123); (ii) a doping-dependent shift of the chemical potential $\mu$ (measured with respect to the aforementioned models);  (ii) a $d$-wave pairing gap $\D_k = \D_0/2\big(\cos(k_x)-\cos(k_y)\big)$; (iv) an inverse quasiparticle lifetime $\Gamma_k$, describing the inelastic scattering of quasiparticles. Microscopically, $\Gamma_k$ is due to electron-electron interactions, but it can also be conveniently described as the inelastic scattering of quasiparticles on dynamic charge and spin fluctuations  \cite{vojta06}. At zero temperature, $\Gamma_{k}$ vanishes for quasiparticles with $\e_k=\mu$, but it is generically finite and positive elsewhere. Because, as we will show, STM and REXS signals are dominated by the scattering of quasiparticles with specific momentum (antinodal quasiparticles), these experiments are well described by the simplifying assumption of a constant $\Gamma_k\equiv\Gamma$  \cite{dynes78} (see also \appendix \ref{sec:antinodal}, where we consider the effects of an anisotropic scattering rate).

STM probes the differential conductance $g(r,V) \equiv dI(r,V)/dV = {\rm Im}\left[{\mathcal G}(r,V)\right]$. In the presence of weak time-independent scatterers its Fourier transform is given by \cite{podolsky03,lee03,polkovnikov03,hirschfeld04,markievicz04,nowadnick12},  ${\mathcal G}(k,k',V)= G(k,V)\delta_{k,k'} + G(k,V)T(k,k')G(k',V)$, where $T(k,k')$ describes the scattering of quasiparticles from momentum $k$ to momentum $k'$. In the case of long-range-ordered waves the elements of $T(k,k')$ are sharply peaked around the ordering wavevector $q=k-k'$. In contrast, for local impurities the scattering amplitude does not depend on the momentum difference and, for $q\neq0$, we have (see also \appendix \ref{sec:gmap})
\be g(q,V) = e^{i q r_0} \sum_{k-k'=q} {\rm Im}\Big[G(k,V)\Big(T_{k}+T_{k'}\Big)G(k',V)\Big]\;.~\label{eq:rhoqw}\ee 
Here $r_0$ is the position of the impurity and $T_k$ is a $k$-dependent $2\times2$ matrix describing the scattering process. 

Theoretically, the nature of the scattering matrix $T_k$ can be deduced from the phase of the Fourier-transformed STM signal  \cite{podolsky03,zhang04}. However, in the raw data the phase also depends on the random position of the impurities $r_0$ and may vary across the sample. To overcome this problem we introduce a new method of analyzing STM spectra, relying on the comparison of the Fourier components at different voltages. In the Methods Section we demonstrate that, for  wavevectors $\vec{q}$ along the Cu-O axis, $g(q,V)$ is mostly symmetric with respect to $V\to-V$, indicating that the elastic scattering at these wavevectors is mainly due to local modulations of the pairing gap  \cite{kapitulnik03,podolsky03,zhang04}.  As we will show, these local impurities can induce Friedel oscillations observed as non-dispersive peaks in the STM signal. The same model also describes localized magnetic vortices (where the pairing amplitude is strongly suppressed), in whose vicinity the incommensurate order was first observed  \cite{hoffman02B}. 

By comparing the intensity of the predicted signal, Eq.~(\ref{eq:rhoqw}), with the experimental observations at an arbitrary wavevector we are able to uniquely determine the pairing gap $\Delta_0$, the quasiparticle lifetime $\Gamma$, and chemical potential $\mu$ throughout the whole superconducting dome (see Methods Section and Table \ref{table1}).  We find that both $\Delta_0$ and $\Gamma$ increase with underdoping, i.e. as approaching the antiferromagnetic insulating phase, in agreement with previous theoretical calculations (FLEX approximation  \cite{dahm95,pao95} and functional RG  \cite{ossadnik08}) and experimental observations (magneto-resistance of ${\rm Tl_2Ba_2CuO_{6+\delta}}$  \cite{hussey06} and STM of Bi2212  \cite{alldredge08}).

\begin{figure}[p]
\begin{tabular}{c c c}
\includegraphics[scale=0.7]{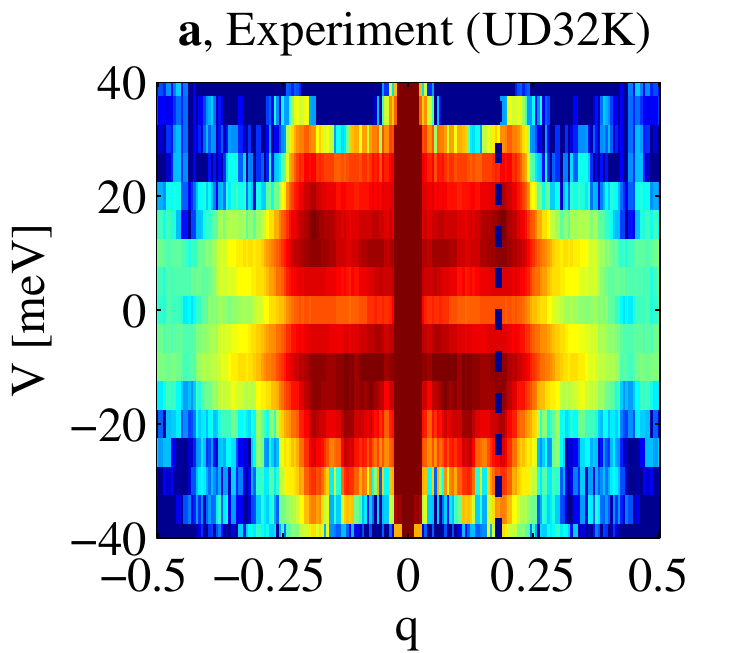} &%analysis_v6_fig2	
\includegraphics[scale=0.7]{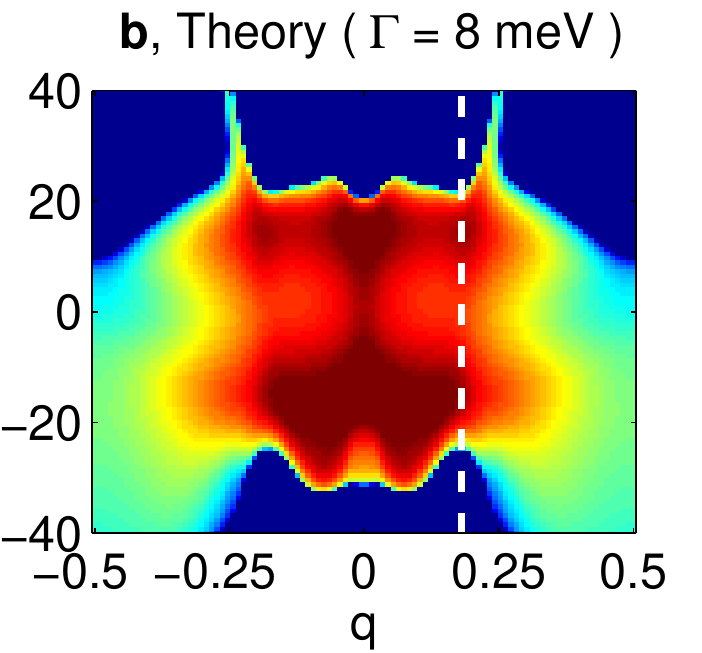} &%singleband_v6_fig2
\includegraphics[scale=0.7]{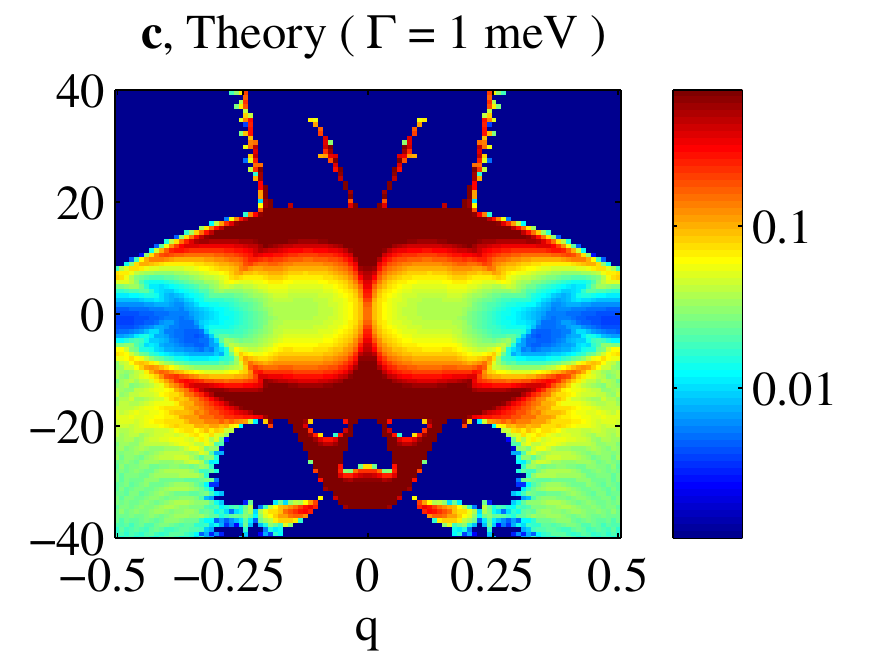} %singleband_v6_fig2
\end{tabular}
\includegraphics[scale=0.53]{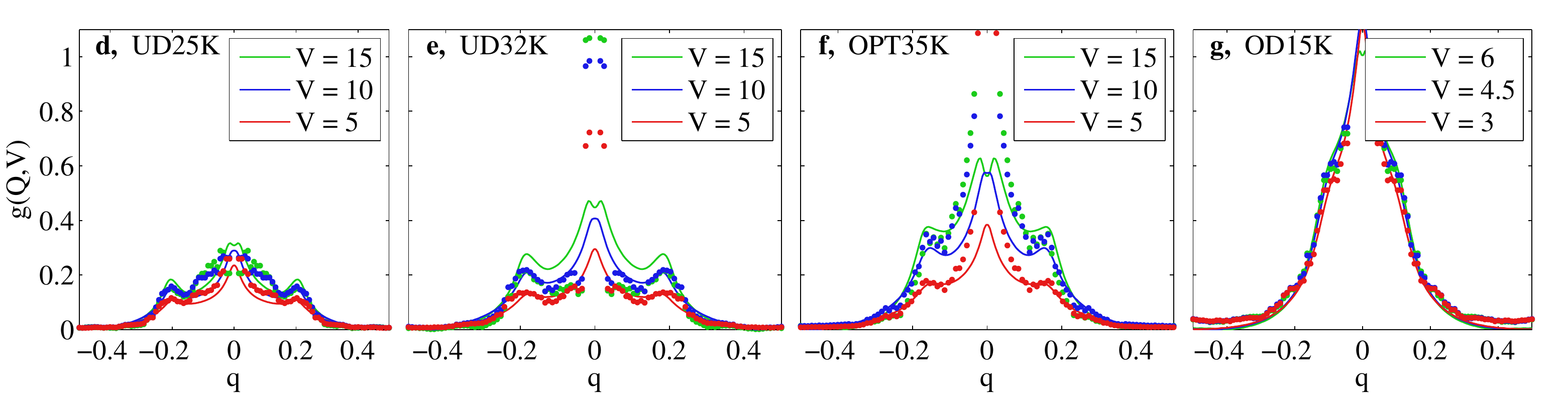}
\caption{{\bf Wavevector and voltage dependence of STM signal at $\bf \vec q=(q,0)\times 2\pi$}. {\bf a-b}, Experiment and theory relevant to an underdoped sample of Pb-Bi2201 (UD32K, see Table \ref{table1}). A non-dispersive peak is observed at $q\approx0.2$ for $5{\rm{meV}}\lesssim{V}\lesssim20{\rm{meV}}$. The dashed lines are guides for the eyes. The intense line around $q=0$ in the experimental result is due to the homogeneous component which was neglected in the theoretical calculation (see \appendix \ref{sec:homo}) . {\bf c}, Same as {\bf b} with a smaller $\Gamma$, showing dispersive features at small voltages ($|V|<20$meV) and non-dispersive features at large positive voltages ($V>20$meV).  {\bf d-g}, Comparison between theory (lines) and experiment (dots) for four  samples of Pb-Bi2201. The theoretical curves were computed from Eq.~(\ref{eq:rhoqw}) using the values of $\Delta_0$, $\Gamma$, and $\mu$ obtained from the analysis of a single wavevector (see Methods section) and without any further rescaling. The voltage is expressed in  meV.}\label{fig:25_comp}%singleband_v7_fig1
\end{figure}

Having extracted the three phenomenological parameters from a single wavevector, we can compute the STM signal at all wavevectors {\it without any additional fitting parameter}. Figure~\ref{fig:25_comp}{\bf a} shows our theoretical predictions for wavevectors aligned along the Cu-O axis, $\vec q=(q,0)\times 2\pi$. Our calculations clearly indicate a non-dispersive peak at wavevector $q\approx0.2\times2\pi$ for voltages $5{\rm{meV}}\lesssim{V}\lesssim20{\rm{meV}}$, which was observed in experiments (see Ref.~\cite{hudson08} and Fig.~\ref{fig:25_comp}{\bf b}).  Figs.~\ref{fig:25_comp}{\bf d-g} demonstrate that our model quantitatively predicts the wavelength and the width of the incommensurate peaks at all dopings. Remarkably, our model includes only scattering from local impurities, without any long-range density or pairing waves. The peaks observed in the STM experiments result from an enhanced scattering of antinodal quasiparticles (see also \appendix \ref{sec:antinodal}). Because the quasiparticles have a finite lifetime, their energy does not need to be exactly conserved during a scattering event. The scattering at wavevectors connecting antinodal quasiparticles is then strongly enhanced at all voltages, giving rise to broad non-dispersive peaks in the STM signal.

To highlight the effects of a finite quasiparticle lifetime, we repeat the above calculations for a smaller value of $\Gamma$ (Fig.~\ref{fig:25_comp}{\bf c}). The predicted signal displays dispersive features for $V<\Delta_0$ and sharp non-dispersive features for $V>\Delta_0$, in agreement with low-temperature STM measurements of Bi2212 (See Ref.~\cite{fujita11} for a review). Our theoretical calculations predict the appearance of an intermediate regime in which both types of peaks disappear and are substituted by broad non-dispersive peaks. As shown in \appendix \ref{sec:twoscales}, the upper and lower boundaries of this intermediate regime are proportional to $\D_0-\Gamma$ and  $\D_0+\Gamma$ respectively. With increasing underdoping, both $\Delta_0$ and $\Gamma$ become larger  \cite{alldredge08} and the two boundaries move apart: the lower boundary is approximately constant, while the upper one shifts to higher energies. (See also \appendix \ref{sec:homo} for the appearance of two energy scales in the real-space spectra.) Accordingly, at higher temperatures \cite{yazdani04} and at lower doping values \cite{yazdani13B}, the non-dispersive peaks were observed to extend down to zero voltage.  An analogous interplay between dispersive and non-dispersive peaks was observed in the autocorrelation analysis of ARPES data \cite{McElroy06,Chatterjee06}. We propose that $\Gamma$ may play the role of the second energy scale detected in the pseudogap phase, approached with increasing underdoping and/or temperature. 

Remarkably, the non-dispersive peaks were observed to persist to the temperature $T^*\gg T_c$  \cite{yazdani10}, marking the transition from the pseudogap phase to the normal phase. In our model the intensity of the non-dispersive peaks is roughly proportional to $\Delta_0$, indicating that a local superconducting gap may be present in the pseudogap phase, even though its long range coherence is already suppressed. As pointed out by Fischer \etal  \cite{Fischer2007}, this argument is consistent with the observation that the ratio $T^*/\Delta_0$ is approximately constant in all cuprates.  Experimental evidence for superconducting fluctuations well above T$_c$	 also comes from recent $\mu$SR data by Mahyari \etal  \cite{sonier13}. In \appendix \ref{sec:ARPES} we show that this model is consistent with ARPES measurements as well \cite{kanigel07,dessau12}.

\begin{figure}[p]
{\bf Theory}
\includegraphics[scale=1.02]{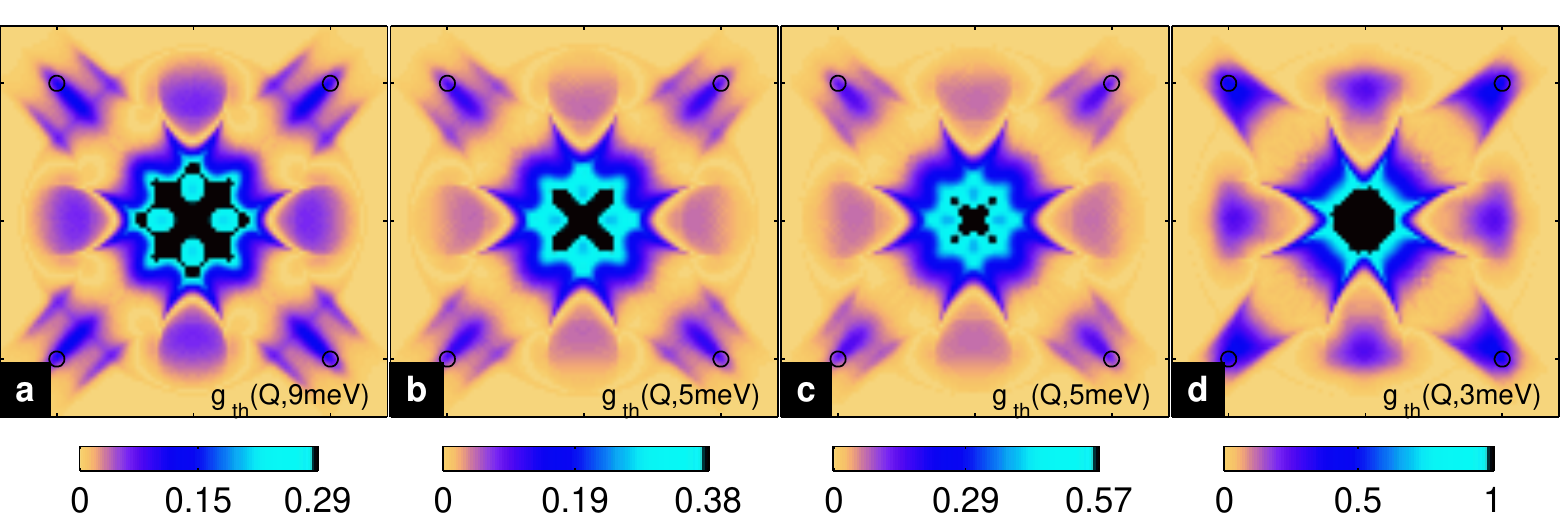}\\%singleband_v7_fig7ter
{\bf Experiment (Pb-Bi2201 \cite{yanghe13})}
\includegraphics[scale=0.29]{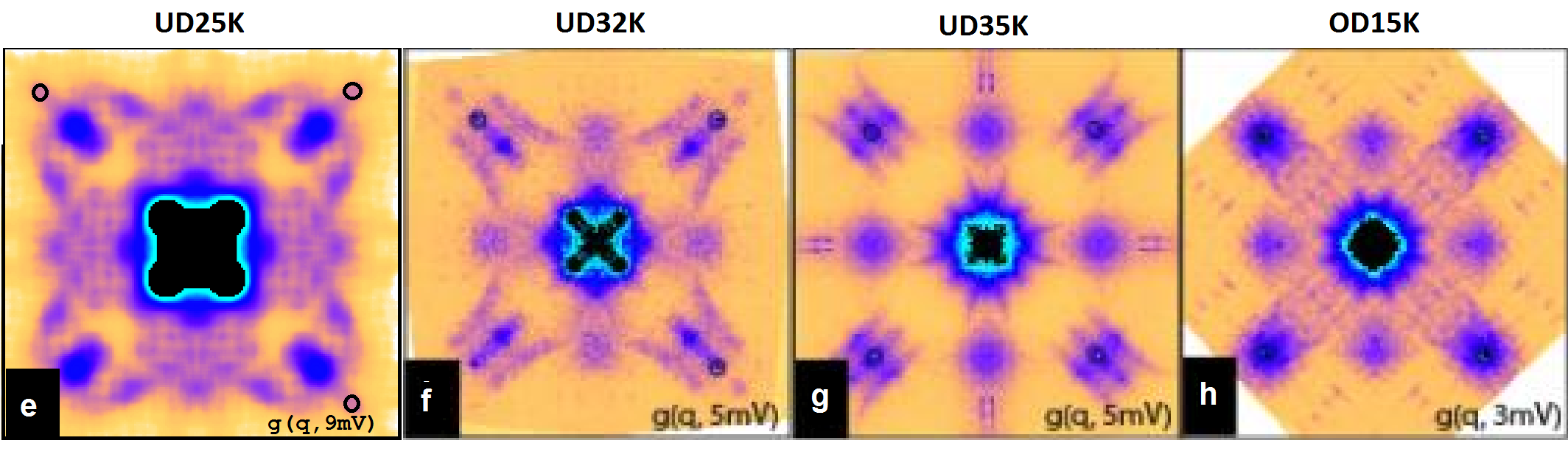}%YangHetal
\caption{{\bf Wavevector dependence of the STM signal at a fixed voltage}. {\bf a-d}, Theoretical predictions, Eq.~(\ref{eq:rhoqw2}), using the physical parameters listed in Table \ref{table1}. {\bf e-g}, Experimental measurements of 4 samples of Pb-Bi2201, reproduced from Ref.~\cite{yanghe13}. In all subplots the small black circles are aid for the eyes and indicate the position of the Bragg peaks $\vec q = (\pm1,0)\times 2\pi$ and $\vec q = (0,\pm1)\times 2\pi$. }
\label{fig:7_comp}
\end{figure}

Up to this point, we considered the STM signal along a specific momentum cut (parallel to the Cu-O axis) only. In order to reproduce the full two-dimensional $(q_x,q_y)$ dependence, we need to take into account two additional factors. First, local modulations of the gap do not scatter quasiparticles at 
 $\vec q_{\pi,\pi}=(0.5,0.5)\times 2\pi$ because at this wavevector the integrand of  Eq.~(\ref{eq:rhoqw}) is identically zero (due to the $p$-wave symmetry of the gap, $T_{k} +  T_{k+q_{\pi,\pi}}\sim \Delta_{k}+\Delta_{k+q_{\pi,\pi}}=0$). In contrast, the experimental signal shows a broad peak around this wavevector. As shown in \appendix \ref{sec:Qpipi}, the peak at $\vec q_{\pi,\pi}$ is due to local modulations of the chemical potential, which coexist with the local modulations of the gap. Along the Cu-O axis, the coherence factors appearing in Eq.~(\ref{eq:rhoqw}) significantly suppress the scattering from modulations of the chemical potential, making the modulations of the gap dominant (see \appendix \ref{sec:antinodal}). To explain the full range of STM results at all wavevectors we need to include both sources of disorder: the experiments are best reproduced by adding modulations of the chemical potential and of the gap with the same amplitude and the same phase. Physically, this implies that the local modulations of $\mu$ and $\D_0$ have a common origin, probably related to the increased density of holes around the dopants. This finding is in agreement with Ref.~\cite{hudson08} who observed a positive correlation between the gap and the wavelength of the incommensurate modulations (which is set by the chemical potential).

Second, Eq.~(\ref{eq:rhoqw}) refers to a lattice model and predicts a signal that is periodic under $\vec q\to \vec q+(n,m)\times2\pi$, where $n$ and $m$ are integers. In contrast, the experimental signal globally decreases as function of $q$. To explain this effect we need to take into account the overlap function $\psi(r)$, describing the tunneling amplitude of quasiparticles from the tip to the sample. This leads to a modified version of Eq.~(\ref{eq:rhoqw}), which reads  \cite{podolsky03,yazadani13}:
\be g'(q,V) \equiv e^{i q r_0} \sum_k~{\rm Im}\Big[ \psi(k+q) G(\w=V,k)\Big(T_k+T_{k+q}\Big)G(\w=V,k+q) \psi^*(k)\Big]\label{eq:rhoqw2}\, \ee
where $\psi(k) = \int d^2r ~e^{ikr}\psi(r)$. In our calculations we assume a Gaussian overlap function $\psi(r) = e^{-r^2/2\delta r^2}$, or $\psi(k)=e^{-\delta r^2 k^2/2a^2}$, where $a$ is the lattice constant. The single fitting parameter $\delta r = 0.55a$ is phenomenologically determined by the ratio between the Fourier components at small and large wavevectors, and allows us to reproduce the experimental data for all four samples, as shown in Fig.~\ref{fig:7_comp}.

We now use our understanding of the quasiparticle Green's function to analyze X-ray scattering. In resonant X-ray scattering (REXS) electrons are transferred to the vicinity of the Fermi surface from a core hole, characterized by an energy $E_h$ and an inverse lifetime $\Gamma_h$. As recently shown in Refs.~ \cite{abbamonte12,benjamin13}, the response to REXS is given by the convolution of $g(q,\omega)$ with the response function of the core hole, $1/(\omega-E_h+i\Gamma_h)$. 
%This expression is obtained by neglecting the electric potential of the core hole  \cite{}. 
At zero temperature, the intensity of the REXS signal is then given by
\be
I_{\rm REXS}(q,E) = \left| \int_0^\infty d\w' \frac{g(q,\omega')}{E-\omega'-E_h + i\Gamma_h}\right|^2\label{eq:REXS}
\ee
Here the integral from 0 to $+\infty$ indicates that the X-ray beam can only create electron excitations above the Fermi surface (with positive energy). Sharp peaks in these experiments are considered ``smoking-gun'' evidence of competing charge order.

\begin{figure}[p]

%{\bf Theory}\\
\begin{tabular}{c c}
\includegraphics[scale=0.8]{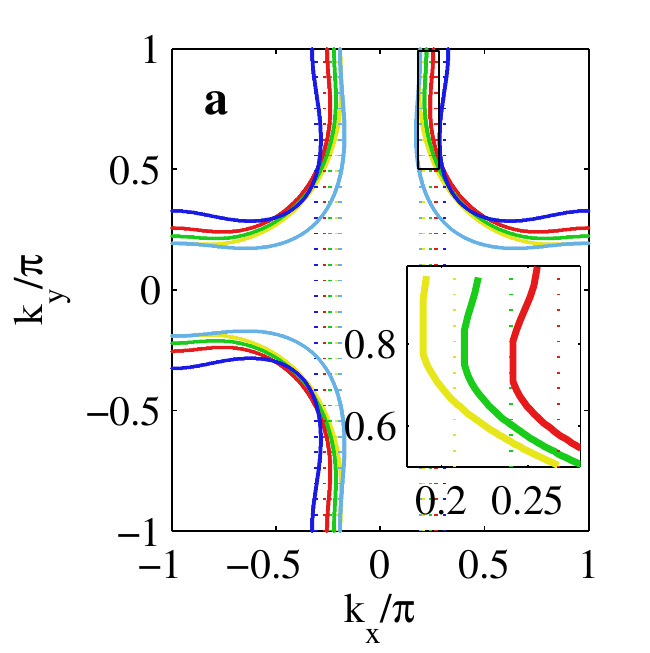}&
\includegraphics[scale=0.8]{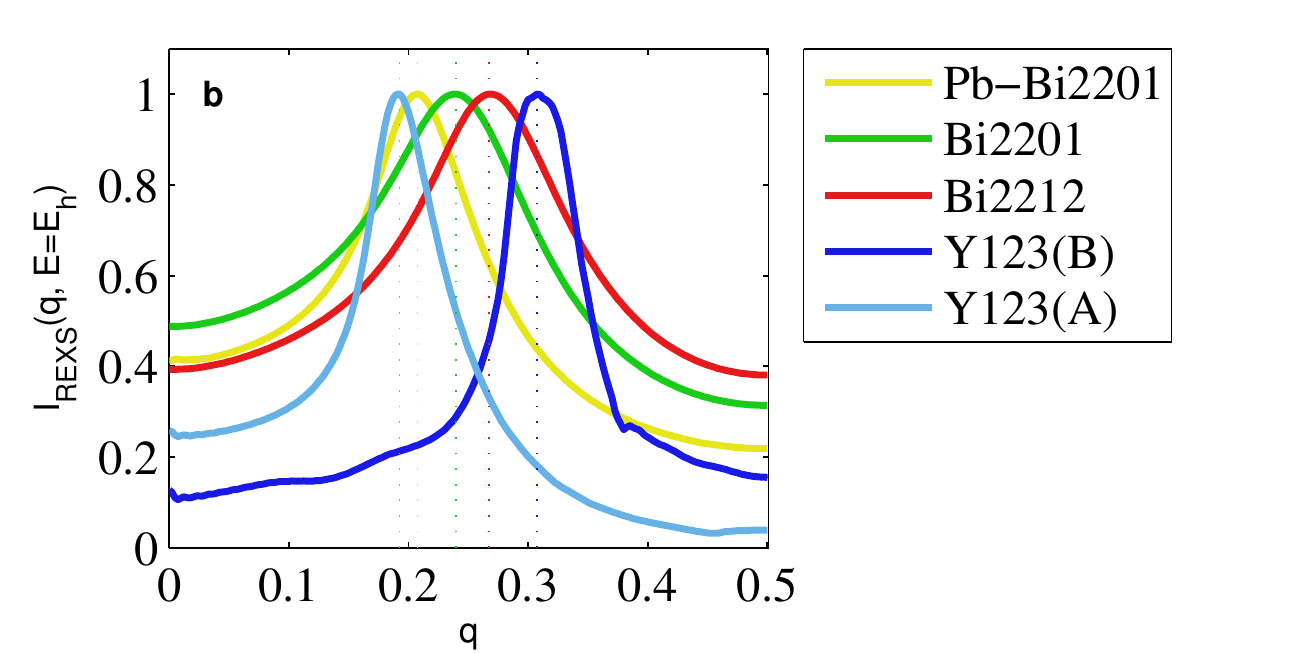}\\
\includegraphics[scale=0.7]{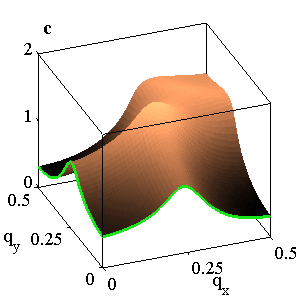} &
%\begin{tabular}{c}
%{\bf Experiment (Y123 \cite{ghiringhelli12})} \\
\includegraphics[scale=0.8]{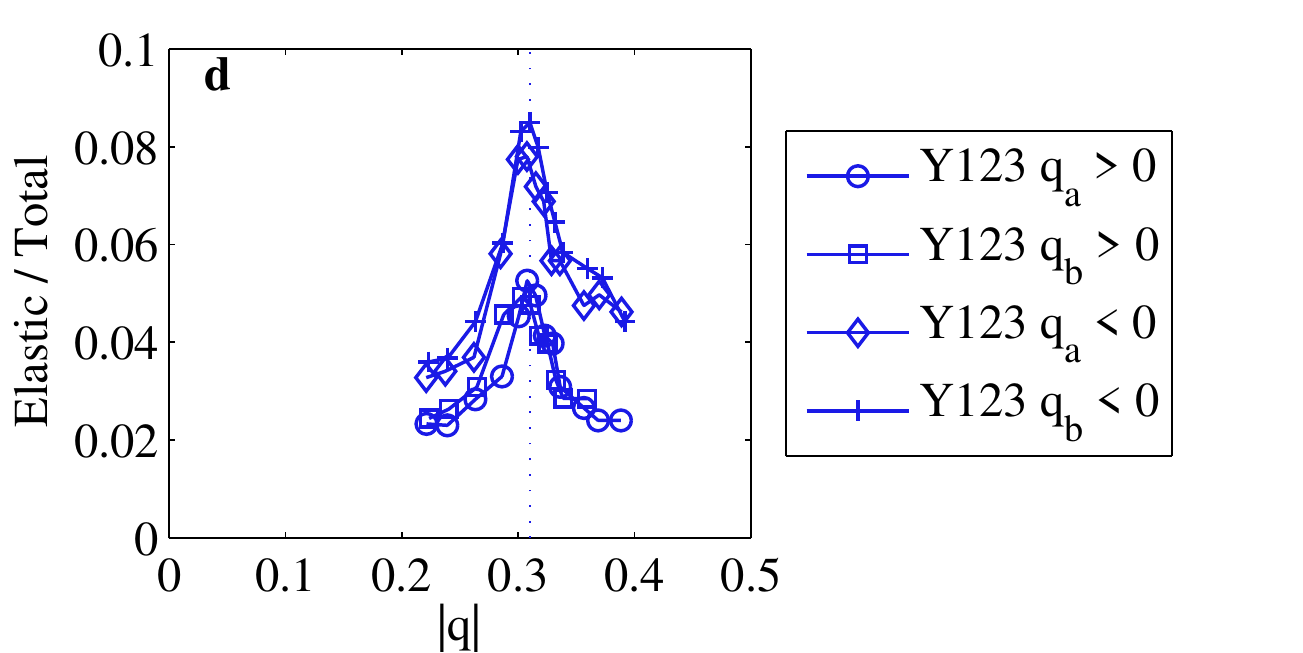}
%\end{tabular}
\end{tabular}
\caption{{\bf Fermi surface and REXS signal}. {\bf a,} Fermi surfaces obtained from the phenomenological band structures of Refs.~\cite{campuzano95,shen98}. The inset presents a detailed view of the antinodal region. {\bf b}, Theoretical predictions of the REXS signal for $\mathbf{\vec q=(q,0)\times 2\pi}$, obtained using Eq.s (\ref{eq:rhoqw}) and (\ref{eq:REXS}). The physical parameters correspond to underdoped samples of five materials: Pb-Bi2201  $(p=0.16)$ -- see Table \ref{table1}. Bi2201, $(p=0.11)$; Bi2212, $(p=0.04)$; Y123 (Bonding (B) and Antibonding (A)) $(p=(p_B+p_A)/2=0.12)$ -- with $\Delta_0=40-60$meV, $\Gamma=10-20$meV. Each curve is normalized to its maximal value. The position and width of the peak is determined by the quasi-nested shape of the band structure. In subplot {\bf a} and {\bf b} the vertical dotted lines are guides for the eyes and indicate the wavevector corresponding to the maximal intensity of the REXS signal, which is found to be $\sim10\%$ larger than the distance between the antinodes.  {\bf c}, Theoretical predictions of the REXS signal for Bi2212 $(p=0.11)$ as a function of the two-dimensional wavevector $\vec q =(q_x, q_y)\times 2\pi$. A pronounced peak at $(q_x,q_y)\approx(0.25,0.25)$ is predicted. {\bf d}, Experimental measurements of underdoped samples of Y123 ($T_c=61$K) reproduced from Refs.~\cite{ghiringhelli12}. The subindices ``a'' and ``b'' refer to the two principal axis of the CuO$_2$ plane. See also \appendix \ref{sec:REXS} for a quantitative comparison between theory and experiment. }
\label{fig:REXS}
\end{figure}

Combining the above calculations of the STM signal $g(q,V)$ with Eq.~(\ref{eq:REXS}), we predict the REXS signal in underdoped samples of Pb-Bi2201 (p=0.16), Bi2201 (p=0.11), and Bi2212 (p=0.04) and find a pronounced peak even in the presence of local scatterers only (Fig.~\ref{fig:REXS}{\bf b}). In analogy to the STM signal described above, this peak is due to enhanced coherence factors in the antinodal regions, amplified by the nearly-nested Fermi surface of cuprates. As shown in Eq.~(\ref{eq:rhoqw}), the predicted signal is determined by a weighted sum over all momenta, leading to a maximal intensity at a wavevector that is approximately $10\%$ larger than the antinodal nesting wavevector (see inset of Fig.~\ref{fig:REXS}{\bf a}). The width of the peak (full-width half-maximum    $\delta q \approx 0.1$) corresponds to an estimated correlation length of approximately $10$ atoms, or $\xi \approx 40\AA$, and is in quantitative agreement with recent measurements \cite{damascelli13,yazdani13B}. A similar effect was observed in X-ray scattering experiments of Y123 \cite{ghiringhelli12,chang12}. Our calculations for the bonding (B) band of Y123 exactly reproduce the wavevector $q\approx0.31$ of the observed signal. The computed width $\delta q \approx 0.07$ is  larger than the one extracted from the experiments ($\delta q = 0.04\pm0.01$ corresponding to $\xi= 100\pm20\AA$ \cite{ghiringhelli12,chang12}). As shown in \appendix \ref{sec:REXS}, $\delta q$ strongly depends on the details of the band structure in the antinodal region, which are generically hard to determine from ARPES measurements. In Y123 the precision of these measurements is further impaired by the polarity of the unit cell and the presence of CuO chains \cite{pasani10}. Both effects are absent in Bi2201 and Bi2212, where we expect our predictions to have a better accuracy. In Fig.~\ref{fig:REXS}{\bf c} we predict the two-dimensional dependence of the REXS signal for a sample of Bi2201 with hole doping $p=0.11$. We predict that, in addition to the peak at $\vec q\approx (0.25,0)\times 2\pi$, a pronounced peak at $\vec q\approx (0.25,0.25)\times 2\pi$ should be observed, highlighting the checkerboard nature of Friedel oscillations (see also \appendix \ref{sec:REXS}). We emphasize that Eq.~(\ref{eq:REXS}) has only two free parameters, $E_h=931.5$eV and $\Gamma_h=400$meV, which can be inferred from the position and width of the x-ray absorption (XAS) peak \cite{ghiringhelli12}, while all other parameters are fixed from ARPES and/or STM measurements. The REXS signal shown in Fig.~\ref{fig:REXS}{\bf b-c} are therefore model-independent consequences of previously-measured quantities. 

To summarize, in this paper we studied the effect of a finite lifetime of quasiparticles on STM and REXS, and provided a minimal framework to quantitatively describe these experiments. We showed that the inverse lifetime $\Gamma$ can play the role of a second energy scale detected by different observations. In particular, we demonstrated that the incommensurate checkerboard short-range order observed in superconducting cuprates Pb-Bi2201, Bi2201, Bi2212, and Y123 can be quantitatively described as the scattering of short-lived quasiparticles on local impurities, in close analogy to Friedel oscillations in a Fermi liquid. This is in contrast to the strongly-coupled unidirectional stripes revealed in the normal phase of other cuprates, whose long correlation length indicates the onset of a true long-range order. In the present analysis we employed a perturbative expansion in the disorder strength: from its success we infer that the effects of the disorder are weak and should not strongly affect $\Gamma$. We therefore suggest that the finite lifetime of quasiparticles at low temperatures is due to inelastic processes, possibly enhanced by the interplay between charge and spin degrees of freedom characteristic of cuprates. Both STM and ARPES \cite{kanigel06,vishik12} measurements clearly indicate that $\Gamma$ rapidly increases with temperature and underdoping. This finding suggests that a finite quasiparticle lifetime may have a significant role in the determination of the dome shape of the critical temperature of cuprates (see also \appendix \ref{sec:phasediagram}). In particular, it is possible that the suppression of the critical temperature in underdoped samples could be due to a decreased quasiparticle lifetime.
 
%MOMENTUM DEPENDENCE OF GAMMA

%TEMPERATURE DEPENDANCE OF GAMMA

%Before concluding, we mention two fundamental points, which deserve further investigation. First, experiments  \cite{davis08_mott,yanghe13} show that, in the underdoped samples, the contributions from the antinodal quasiparticles terminate at the anti-ferromagnetic boundary, perhaps due to the vicinity to the Mott insulating phase. Second, the experimental signal is asymmetric with respect to $x\leftrightarrow y$, possibly due to the spontaneous breaking of this symmetry \cite{davis10,schmidt11,fujita11}. In the literature, it has been shown that both effects can be {\it qualitatively} reproduced by considering specific types of scattering matrices  \cite{vishik09,nowadnick12} and overlap functions  \cite{yazdani13}. Whether this route can provide an explanation to all the observed features is a question that can be answered through the quantitative methods presented here.

%%%%%%%%%%%%%%%%%%%%%%%%%%%%%%%%%%%%%%%%
%

\section*{Methods}

In this section we show how to identify the main source of scattering and the three phenomenological parameters ($\Delta_0$, $\Gamma$, and $\mu$) by comparing Eq.~(\ref{eq:rhoqw}) with STM experimental data. Our method consists in the analysis of a {\it single} point in momentum space. As shown in the text, the full momentum dependence can be predicted without any further fitting parameter. 

The theoretical model presented in the text depends on a yet-to-be-determined $2\times2$ matrix, $T_k$, describing the effects of a local scatterer. The two main sources of scattering a local modulations of the pairing gap and of the chemical potential. Clearly, the former are diagonal in Nambu space and $k$ independent, while the latter are off-diagonal and possess a $d$-wave symmetry. For completeness, we consider here four distinct scattering operators:
\be
T^{(1)}_k = \left(\ba{c c}1 & 0\\ 0 & -1\ea\right),\;
T^{(2)}_k = \left(\ba{c c}d_k & 0\\ 0 & -d_k\ea\right),\;
T^{(3)}_k = \left(\ba{c c}0 & 1\\ 1 & 0\ea\right),\;
T^{(4)}_k = \left(\ba{c c}0 &d_k \\ d_k & 0\ea\right)\;,
\label{eq:T1234}
\ee
where $d_k=\cos(k_x)-\cos(k_y)$ is a $d$-wave function. The operators $T^{(1)}_k$ and $T^{(4)}_k$ correspond respectively to local modulation of the chemical potential and of the pairing gap. As explained in the text, the latter dominates the experimental signal on the Cu-O axis and the former around $\vec{q}_{\pi,\pi}$.

\begin{figure}[b]
\centering
{\bf Theory}
\includegraphics[scale=0.65]{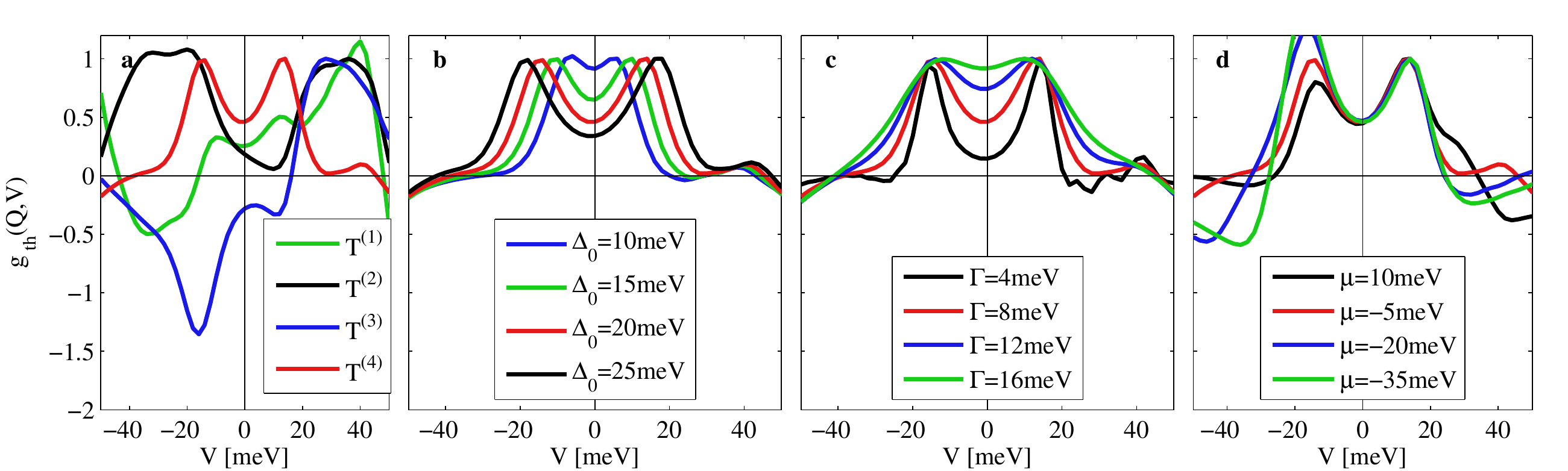}
{\bf Experiment (Pb-Bi2201 -- UD32K)}
\includegraphics[scale=0.75]{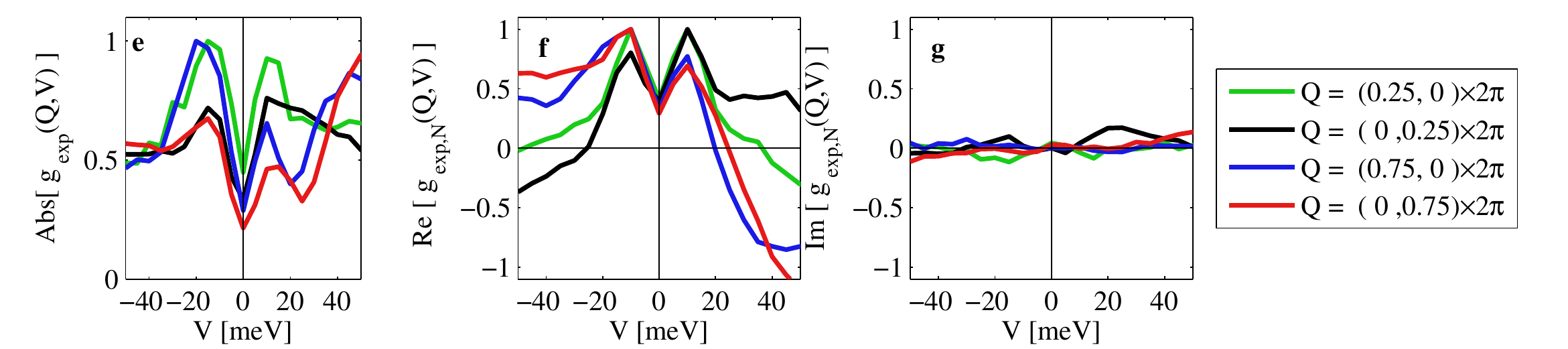}
\caption{{\bf Voltage dependence of the STM signal at fixed wavevector $\bf \vec q=(0.25,0)\times 2\pi$}. {\bf a-d}, Theoretical predictions $g_{th}(q,V)\equiv e^{-i q r_0}g(q,V)$ (see Eq.~(\ref{eq:rhoqw})), obtained by summing over up to $N=500\times500$ discrete $k$-points. In each subplot a single parameter is varied: {\bf a}, Different types of scattering operators (see Eq.~(\ref{eq:T1234})); {\bf b}, Different values of the gap (in meV); {\bf c}, Different values of the quasiparticle lifetime (in meV); {\bf d}, Different values of the chemical potential (in meV). The red curves are identical throughout the plots and refer to the values of $\a=4$, $\D_0=20$meV, $\Gamma=8$meV, $\mu=-5$meV, which best reproduce the experimental measurements.~{\bf e-g}, Experimental measurement for an underdoped sample of Pb-Bi2201 with critical temperature $T=32K$ (UD32K), at four equivalent wavevectors. {\bf e}, Absolute value after smoothing $|g_{\rm exp}(q,V)| = \sum_{p, |p-q|<\d q} |g(p,V)|$; {\bf f-g}, Real and imaginary part of the normalized experimental signal $g_{\rm exp,N}(p,V) = \sum_{|p-q|<\d q} g(p,V)e^{-i\phi(q)}$, where $\phi(q)$ is defined in the text. In all plots each curve is normalized independently. }\label{fig:1_comp}
\end{figure}

In Fig.~\ref{fig:1_comp} we isolate the effects of the free parameters of our theory by varying each of them independently. In Fig.~\ref{fig:1_comp}{\bf a} we vary the scattering matrix $T_k$ among the four options of Eq.~(\ref{eq:T1234}), and observe dramatic effects on the voltage-dependence of the resulting signal. In particular, we observe that the predicted signal is mainly anti-symmetric with respect to $V\to-V$ for modulations of the chemical potential ($T^{(1)}$) and symmetric for modulations of the pairing gap ($T^{(4)}$). Fig.~\ref{fig:1_comp}{\bf b} shows that $\D_0$ controls the position of the peaks. Note however that the maxima are not located at $\pm\D_0$ (as often assumed in the literature), but rather at approximately $\pm0.75\Delta_0$. As explained in \appendix \ref{sec:antinodal}, this is due to the fact that the STM signal is due to quasiparticles in a broad range of momenta (close to the antinodes), whose energy is necessarily smaller than $\Delta_0$. Figure \ref{fig:1_comp}{\bf c} shows that the inverse quasiparticle lifetime $\Gamma$  controls the width of the peaks and the amplitude of the zero-voltage-conductance. Finally, Figure \ref{fig:1_comp}{\bf d} shows that the chemical potential $\mu$  controls the relative intensity of the two peaks. 

We now move to the experimental data. The absolute value of the Fourier transformed signal $|g_{\rm exp}(q,V)|$ is shown in Fig.~\ref{fig:1_comp}{\bf e} at four $q$-points that are equivalent under the lattice symmetry group. As usual, the raw data is smoothed by averaging over a small region in q-space $\delta q\approx0.03\times2\pi$ to increase the signal-to-noise ratio. Note that, for wavevectors inside the Brillouin zone  (green and black curves), the differential conductance is peaked at $V\approx\pm15$meV, while for larger wavevectors (red and blue curves) its maximum is at $V\approx 40$meV. A similar behavior was observed in Ref.~\cite{yazdani10} and to date not explained. As shown in \appendix \ref{sec:normalization}, this effect is probably related to the normalization procedure required to analyze the STM data.

To refine our analysis, we develop a new method that allows us to extract the complex amplitude of the Fourier component of the experimental signal. Here the difficulty is due to the arbitrary phase of the different Fourier components, determined by the location of impurities in the sample (Eq.~\ref{eq:rhoqw}). Due to this random phase, the smoothing techniques presented above cannot be straightforwardly applied. To overcome this problem, we first divide the signal at each wavevector by the corresponding low-voltage phase, $e^{i\phi(q)} = \sum_{|V|<V_{\rm max}} g(q,V)~/~|g(q,V)|$, where $V_{\rm max}=10$meV is an arbitrary cutoff (see also \appendix \ref{sec:Qpipi} for a different choice of $e^{i\phi(q)}$ leading to similar results). This allows us to subsequently average over $\delta q$ and isolate the real and imaginary components of the signal, as shown in Fig.~\ref{fig:1_comp}{\bf e-g}. The imaginary part is small, indicating that the phase of the signal is voltage independent, in agreement with our model of scattering from time-independent perturbations. The real part is analogous to the signal observed in Bi2212  \cite{kapitulnik03} and is now suitable for a direct comparison with the theoretical predictions (Fig.~\ref{fig:1_comp}{\bf a-d}). The experimental signal is symmetric with respect to $V\to-V$, demonstrating that the scattering is dominated by local modulations of the gap ($T^{(4)}$)  \cite{podolsky03,zhang04}. The position of the peaks ($V\approx \pm 15$meV) indicates that the pairing gap is $\Delta_0\approx 20$meV. From the width of the peaks and their relative height, we deduce that $\Gamma\approx 8$meV and $\mu\approx-5$meV. By repeating this approach for the other three samples (not shown here) we obtain the parameters presented in Table \ref{table1}. The values of the gaps obtained by the present analysis coincide with the superconducting gaps found in ARPES measurements on the same material (see Ref.~\cite{shen12} and references therein). Surprisingly, these experiments showed strong deviations of the gap from a simple $d$-wave form, leading to a two-fold-larger gap around the antinodes ($\Delta_{\rm antinode}\approx 40$meV). This larger gap, which may be related to a competing order, is not observed in STM experiments (see Fig.~\ref{fig:25_comp}{\bf a}). This point deserves further investigation.

\begin{table}[p]
\begin{tabular}{|c |c| c |c| c|}
\hline
&  ~~$\Delta_0$~~ & ~~$\Gamma$~~ & ~~$\mu$ (p)~~ \\
\hline
OD15K  & 8 & 6 & -30 (p=0.33)  \\
\hline
OPT35K  & 18 & 7 & -15 (p=0.24)  \\
\hline
UD32K  & 20 & 8 & -5  (p=0.20) \\
\hline
UD25K & 22 & 10 & 5 (p=0.16)\\
\hline
\end{tabular}
\caption{{\bf Phenomenological parameters found by comparing theory and experiment at $\bf \vec q=(0.25,0)\times 2\pi$ for four samples of Pb-Bi2201}. The name of the sample indicates whether the sample is overdoped (OD), optimally doped (OPT), or underdoped (UD) and its critical temperature. The chemical potential is measured with respect to the band structure of Ref.~\cite{campuzano95}.  The concentration of holes (p) is obtained from the Luttinger count (i.e. counting the number of states with $\e_k<\mu$) and found in good agreement with ARPES measurements in the normal phase  \cite{kondo06}. All energies are given in meV and the precision of each entry is of about $10\%$.}
\label{table1}
\end{table}

%WE DON'T MENTION THE SYMMETRIC NORMALIZATION BECAUSE IN FIG.1 IT HAS ALMOST NO EFFECT!
%y relevant to the momenta in the second Brillouin zone show a strong asymmetry between positive and negative voltages, which has been already observed by  \cite{yazdani}, and not understood so far. 
%One possible origin of the asymmetry between positive and negative voltages is related to the ``STS working condition'', in which the current through the tip is normalized at a finite negative voltage $V_{\rm set}$. By definition, this procedure cause the Fourier transformed signal at $V=V_{\rm set}$ to be identically equal to zero and, accordingly, distorts the voltage dependence of the signal. We solve this problem by going back to real space and normalizing the data so that 
%\be I(x,V_{\rm min})-I(x,V_{\rm max}) = \int_{-V_{\rm min}}^{V_{\rm max}} dv~ dI/dV = {\rm const}\label{eq:norm}.~\ee 

\section*{Acknowledgments} We are grateful to Mike Boyer, Kamalesh Chatterjee, Eric Hudson, Jennifer Hoffman, and Doug Wise for giving us access to their unpublished experimental data. We also acknowledge Mehrtash Babadi, Erez Berg, Andrea Damascelli, J. C. S\'eamus Davis, Thierry Giamarchi, Jennifer Hoffman, Yachin Ivry, Amit Kanigel, Malcolm Kennett, Nimrod Moiseyev, Elisabeth Nowadnick,  Daniel Podolsky, Subir Sachdev, Jeff Sonier, and Philipp Strack for many useful discussions.

\newpage

\renewcommand{\theequation}{S\arabic{equation}}   
\setcounter{equation}{0}

\renewcommand{\thefigure}{S\arabic{figure}}   
\setcounter{figure}{0}

\renewcommand{\thesection}{SI-\arabic{section}}   
\setcounter{section}{0}

\renewcommand{\thetable}{S\arabic{table}}   
\setcounter{table}{0}

%\appendix

\def \myparagraph#1 {\section{#1}}

\section*{Supplementary Information (SI)}

\begin{figure}[p]
\centering
{\bf Theory}\\
~~~~~~~~\includegraphics[scale=0.55]{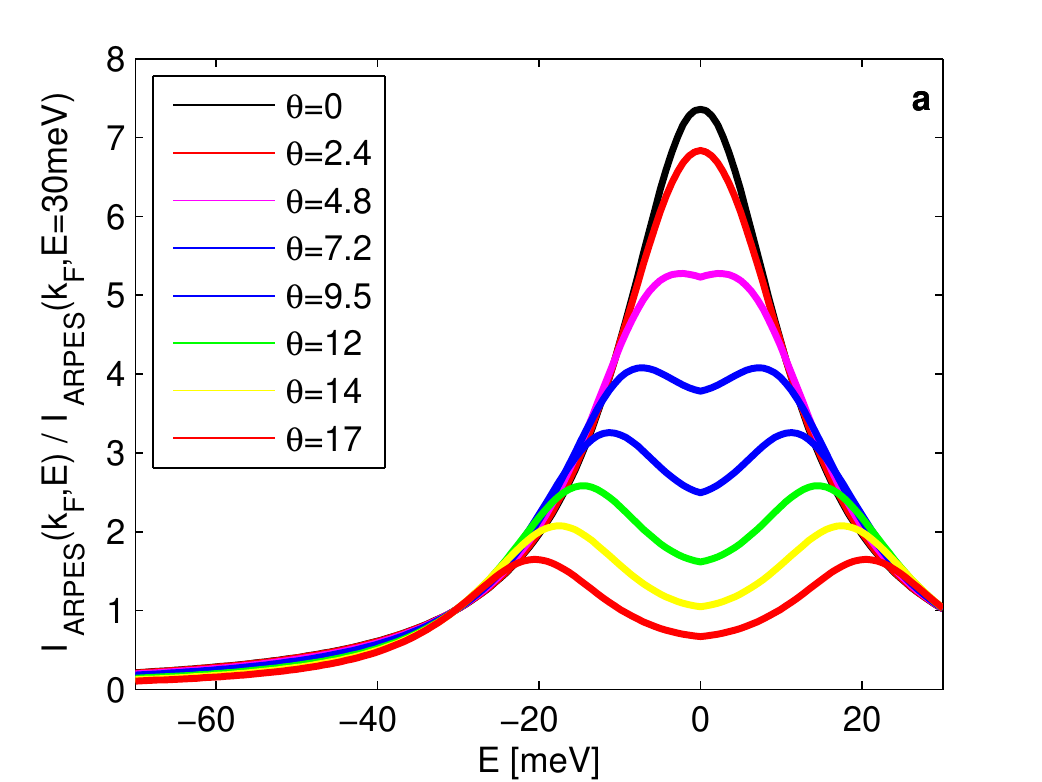} ~~~\includegraphics[scale=0.65]{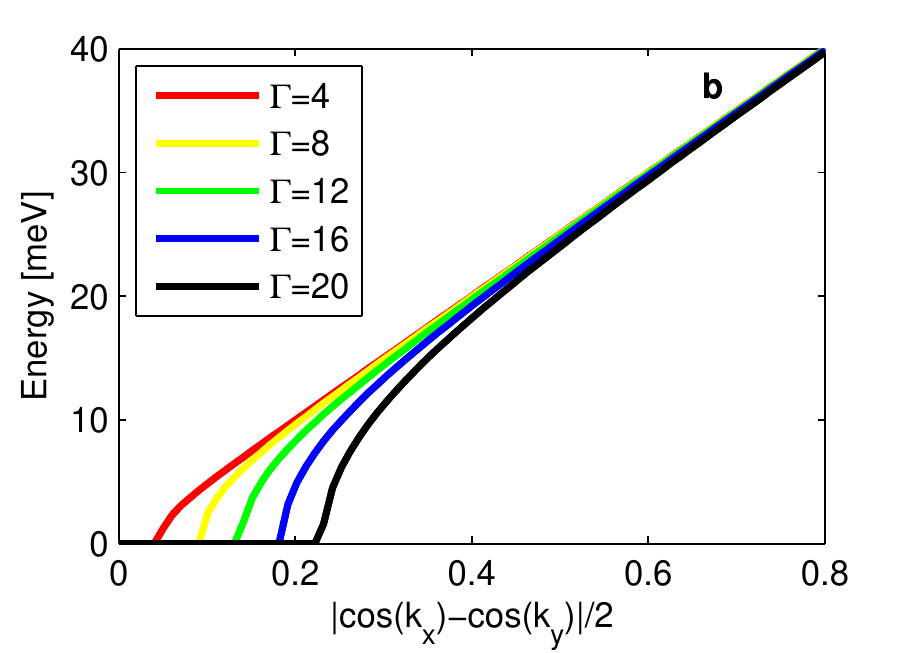}\\
{\bf Experiment}\\
\includegraphics[scale=0.5]{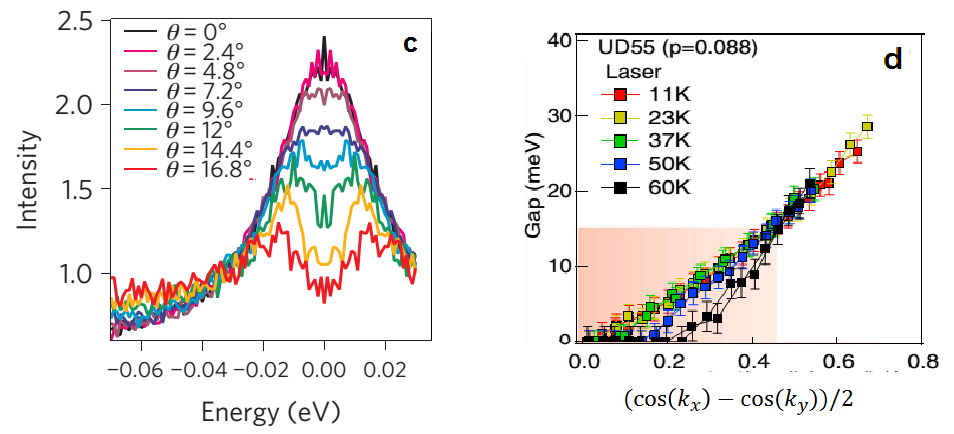}
\caption{{\bf Energy gap on the Fermi surface as detected by ARPES}. {\bf a}, Predicted spectral function, Eq.(\ref{eq:ARPES}), for different points on the Fermi surface. The angle $\theta$ is measured with respect to the nodal points ($\theta=45$ corresponds to the antinodes). The phenomenological parameters are $\Gamma=12$meV, and $\Delta_0=40$meV, and $\mu=20$meV $(p=0.11)$. {\bf b,} Position of the maximal intensity of the ARPES signal on the Fermi surface, Eq.~(\ref{eq:Emax}), as function of $\sin(2\theta)$ for increasing values of $\Gamma$, with $\Delta_0=50$meV, and $\mu=30$meV $(p=0.08)$. {\bf c-d}, Experimental measurements of underdoped samples of Bi2212 from Refs.~\cite{dessau12} (subplot {\bf c}, $T_c=65$K, measured at $75K$) and  \cite{vishik12} (subplot {\bf d}, $T_c=55K$).}\label{fig:ARPES}%singleband_v7_fig1
\end{figure}

\myparagraph{ARPES spectra and Fermi arcs}
\label{sec:ARPES}
In the section we discuss implications of our model for ARPES experiments. Norman \etal \cite{kanigel07} and Reber \etal \cite{dessau12} have already pointed out that a finite quasiparticle lifetime provides a natural explanation for the ARPES spectra, including the emergence of Fermi arcs in underdoped samples. Here we review their arguments and relate them to the Green's function formalism used in this paper. At low temperatures ARPES probes the spectral function, defined as the imaginary part of the diagonal elements of $G(q,\w)$  \cite{ARPES_review}. For momenta on the Fermi surface, $\e_k=\mu$, the (symmetrized) ARPES signal is  then given by
\be I_{\rm ARPES}(k,E) = {\rm Im}\left[\frac{E-i \Gamma}{(E-i\Gamma)^2-(\D_{k}^2)}\right]\label{eq:ARPES} \ee
In Fig.~\ref{fig:ARPES}{\bf a, c} we  directly compare the imaginary part of $G(k,\w)$ with the symmetrized spectrum observed in ARPES experiments and find a very good agreement. Eq.~(\ref{eq:ARPES}) behaves differently depending on the ratio $\Gamma/\D_{k_F}$. For $\Gamma/\Delta_k<\sqrt{3}$, it has two maxima at \be E = \pm\sqrt{2\Delta_{\rm k}\sqrt{\D_{\rm k}^2+\Gamma^2}-\D^2_{\rm k}-\Gamma^2}\label{eq:Emax}\;.\ee In Fig.~\ref{fig:ARPES}{\bf b, d} we show that this expression qualitatively reproduces the evolution of the ``Fermi arcs'', provided that $\Gamma$ is assumed to be temperature dependent. For $\Gamma/\Delta_k>\sqrt{3}$ the same curve has a single maximum at $E=0$. As a consequence, ``Fermi arcs'' are expected to be observed in the vicinity of the nodes for all momenta satisfying $\Delta_k = \Delta_0(\cos(k_x)-\cos(k_y))/2 < \Gamma/\sqrt{3}$. The growth of the Fermi arcs with increasing temperature \cite{kanigel06} and underdoping \cite{vishik12} can be explained in terms of a growth of $\Gamma$, rather than a closing gap. Because ARPES directly probes the nodal quasiparticles, while STM is mostly sensitive to antinodal quasiparticles, a systematic comparison of these two methods on the same materials and temperatures will deliver valuable information about the anisotropy of the inelastic scattering, and help to understand its physical origin.

\myparagraph{Theoretical description of STM measurements}
\label{sec:gmap}
In this Appendix we present the derivation of Fig.~(\ref{eq:rhoqw}), describing the Fourier transformed STM signal induced by a single time-independent impurity. As mentioned in the text, STM measures the differential conductance
\be g(r,V)  \equiv \frac{dI(r,V)}{dV} = {\rm Im}\left[{\mathcal G}(r,r,V)\right]\;, \ee
where ${\mathcal G}(r,r',V)$ is the dressed Green's function including the effects of disorder.  In the case of a time-independent scatterer at position $r_0$, first-order perturbation theory gives:
\be
{\mathcal G}(r,r,V)=G(0,V) + \int dr_1 \int dr_2~G(r-r_1,V)T_0(r_1-r_0,r_2-r_0) G(r_2-r,V)
\ee
Here $G(r-r',V)=G(r,r',V)$ is the translational-invariant bare Green's function (\ref{eq:G0}), which includes the effects of interactions. Introducing its Fourier transform $G(k,V)=\int dr~ e^{i k r}~ G(r, V)$ we obtain
\be
{\mathcal G}(r,r,V)=\sum_k G(k,V) + \sum_{k_1,k_2} G(k_1,V)e^{i(k_1-k_2)(r-r_0)} T_0(k_1,k_2)G(k_2,V)\;,
\ee
where $T_0(k_1,k_2)=\int dr_1\int dr_2~ e^{i k_1 r_1-i k_2 r_2} T_0(r_1,r_2)$. 

For a local impurity $T_0(k_1,k_2)=T_0(k_1)+T_0(k_2)$ and
\be
{\mathcal G}(r,r,V)=\sum_k G(k,V) + \sum_{k_1,k_2} e^{i (k_1-k_2) (r-r_0)} G(k_1,V)\left(T_0(k_1)+T_0(k_2)\right){ G}(k_2,V)
\ee

If both the bare Green's function and the impurity scattering are invariant under inversion symmetry $k \to -k$, only the cosine component contributes to the integral and
\be
{\mathcal G}(r,r,V)=\sum_k G(k,V) + \sum_{k_1,k_2} \cos((k_1-k_2)(r-r_0)) G(k_1,V)\left(T_0(k_1)+T_0(k_2)\right){ G}(k_2,V)\;.
\ee
The Fourier transformed STM signal is then 
\bea 
g(q,V) &=& \int dr ~e^{i q r} ~{\rm Im}\left[{\mathcal G}(r,r,V)\right] \\
&=& \sum_k {\rm Im}\left[G(k,V)\right]\delta_{q,0}
 + \int dr~e^{i q r}
\sum_{k_1,k_2} \cos((k_1-k_2)(r-r_0))
\nn\\ &&\times {\rm Im}\left[G(k_1,V)\left(T_0(k_1)+T_0(k_2)\right){\mathcal G}(k_2,V)\right]
\eea
Using the identity $\int dr~e^{i q r}\cos(q' (r-r_0))=e^{i q r_0}\left(\delta_{q+q'}+\delta_{q-q'}\right)/2$ we find
\be
g(q,V) =\sum_k {\rm Im}\left[G(k,V)\right]\delta_{q,0} +\frac 12 e^{i q r_0}\sum_{k_1-k_2=\pm q}  {\rm Im}\left[G(k_1,V)\left(T_0(k_1)+T_0(k_2)\right)G(k_2,V)\right]\label{eq:rhoqwB}
\ee
Due to the above-mentioned symmetry ($k\leftrightarrow -k$) the contributions from terms with $k-k'=q$ and $k-k'=-q$ are identical, and the finite-$q$ components of Eq.~(\ref{eq:rhoqwB}) are given by Eq.~(\ref{eq:rhoqw}).

\myparagraph{Coherence factors of nodal and antinodal quasiparticles}
\label{sec:antinodal}
In the text we explained that the non-dispersive peaks observed in STM originate from enhanced scattering at the antinodes (see also Ref.~\cite{yanghe13}). This observation is in contradiction with the well-known ``octet model''  \cite{lee03}, which predicts that antinodal quasiparticles should contribute only at a specific voltage, $V\approx \Delta_0$, due to energy conservation. Accordingly, for small voltages $V\ll\Delta_0$ only nodal quasiparticles are expected to contribute. We now show that this picture is dramatically changed when the appropriate matrix elements (``coherence factors'') are taken into account. For convenience, we define the integrand of Eq.~(\ref{eq:rhoqw}) as 
\be S^{(\alpha)}(k,k+q,V) = {\rm Im}\left[G(k,V)\left(T^{(\a)}_{k}+T^{(\a)}_{k+q}\right)G(k+q,V)\right]\;.\label{eq:S}\ee Eq.~(\ref{eq:S}) determines the contribution to the differential conductance originating from the scattering of a quasiparticle from momentum $k$ to momentum $k+q$. The experimental observable $g(q,V)$ is obtained by integrating $S$ over all momenta $k$.

Let us first consider modulations of the chemical potential ($\a=1$), in the limiting case of zero voltage and $\Gamma\to0$. In this limit Eq.~(\ref{eq:S}) simplifies to \be S^{(1)}(k,q,0) = \Gamma\frac{((\e_k-\mu)+(\e_{k+q}-\mu))}{(\D_k^2+(\e_k-\mu)^2)(\D_{k+q}^2+(\e_{k+q}-\mu)^2)}\label{eq:cfactor} \ee
The denominator of Eq.~(\ref{eq:cfactor}) vanishes if both $k$ and $k+q$ correspond to the nodal points (where $\e_k=\e_{k+q}=\mu$ and $\D_k=0$) in agreement with the octet picture. However, precisely at this point the numerator vanishes as well and the contribution to the differential conductance is zero. Because on the two sides of the nodal point Eq.~(\ref{eq:cfactor}) has opposite sign (dependending on whether $\e_k$ is larger or smaller than $\mu$), the integral over $k$ gives an almost-vanishing contribution. In this case, the peak predicted by the octet model is completely washed out.

\begin{figure}[p]
\centering
\includegraphics[scale=0.8]{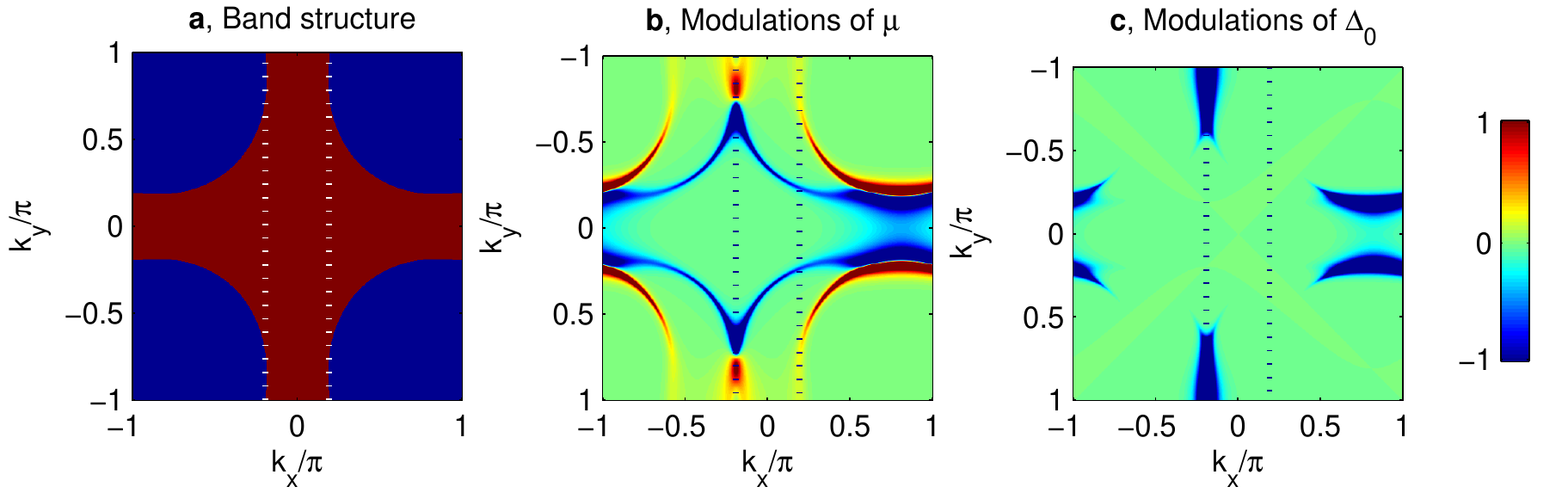}
\caption{{\bf Coherence factors (\ref{eq:S}) as function of momentum $\vec k = (k_x,k_y)$ for Pb-Bi2201 (p=0.16)}. {\bf a} Band structure of this material. We consider the scattering at wavevector connecting two antinodes, $q^*=(0.1924,0)\times 2\pi$ (distance between the dotted lines). {\bf a}, The contributions from modulations of the chemical potential ($S^{(1)}$) change sign as a function of momentum and are strongly suppressed when the sum over $k$ is taken into account. {\bf b}, The contributions from modulations of the pairing gap ($S^{(4)}$) do not change sign and are peaked in a finite region in momentum space around the antinodes.}
\label{fig:8_theory}
\end{figure}

We now consider the role of the coherence factors (\ref{eq:S}) in generating the non-dispersive peak at $q\approx(0.2,0)\times 2\pi$. Fig.~\ref{fig:8_theory}{\bf b} presents a colorplot of $S^{(1)}_(k,k+q,V)$, associated with the modulations of the chemical potential. In agreement with our previous argument, we find that the coherence factors change sign across the Fermi surface, and are strongly suppressed when the sum over all $k$ is taken into account. In contrast, the coherence factors due to modulations of the pairing gap (subplot {\bf c}) do not change sign and therefore dominate the predicted STM signal at this wavevector. It is interesting to compare our results with the octet model. This model predicts contributions from quasiparticles with a specific momentum, given by the intersection between the Fermi surface and the line $k_x=q_x/2$. For $q=q^*$ this momentum is approximately half-way between the nodal and antinodal points. In contrast to the octet model, our approach shows that the STM signal is determined by quasiparticles with a broad range of momenta, close to the antinodal points (blue regions in Fig.~\ref{fig:8_theory}{\bf b}).

\begin{figure}[p]
\centering
\includegraphics[scale=0.7]{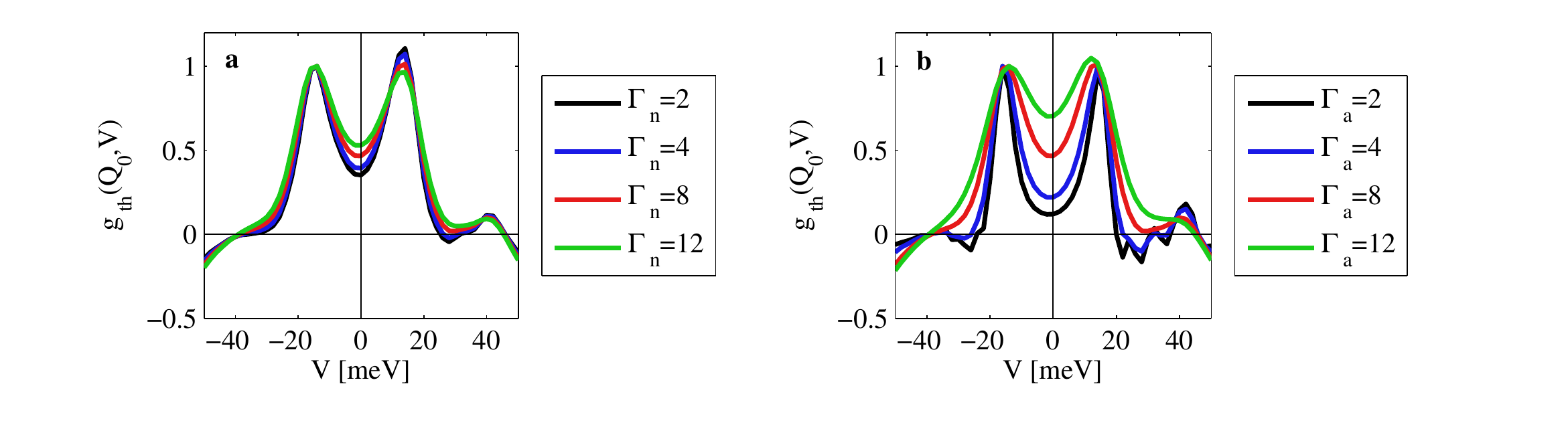}
\caption{{\bf STM signal with a momentum-dependent quasiparticle lifetime.} Same as Fig.~\ref{fig:1_comp}{\bf g}, but with the angular dependent quasiparticle lifetime (\ref{eq:etaq}). The predicted differential conductance depends significantly only on the antinodal quasiparticle scattering}
\label{fig:1bis_theory}
\end{figure}

To further highlight the predominance of antinodal scattering in STM signal we now consider a  momentum-dependent quasiparticle lifetime of the form
\be \Gamma_k = \Gamma_{n} + \frac{|\D_k|}{\D_0}(\Gamma_{a}-\Gamma_{n}) \;,\label{eq:etaq}\ee
where $\Gamma_n$ and $\Gamma_a$ are, respectively, the quasiparticle lifetime at the nodes and at the antinodes. The resulting predictions for the Fourier-transformed STM signal is shown in Fig.~\ref{fig:1bis_theory}. We find that the predicted differential conductance is strongly dependent on $\Gamma_a$ and almost insensitive to $\Gamma_n$. This finding justifies {\it a posteriori} the assumption of a constant $\Gamma_k=\Gamma$, used in our analysis of the STM and REXS signals.
 %From STM measurements alone it is not possible to determine the momentum dependence of $\Gamma$. In contrast ARPES measurements are mostly sensitive to nodal quasiparticles. Combining STM and ARPES measurements for the same materials would allow a clear identification of the momentum-dependent quasiparticle lifetime, allowing a quantitative comparison with magnetotransport measurements  \cite{hussey06}.

\begin{figure}[p]
\centering
\includegraphics[scale=0.65]{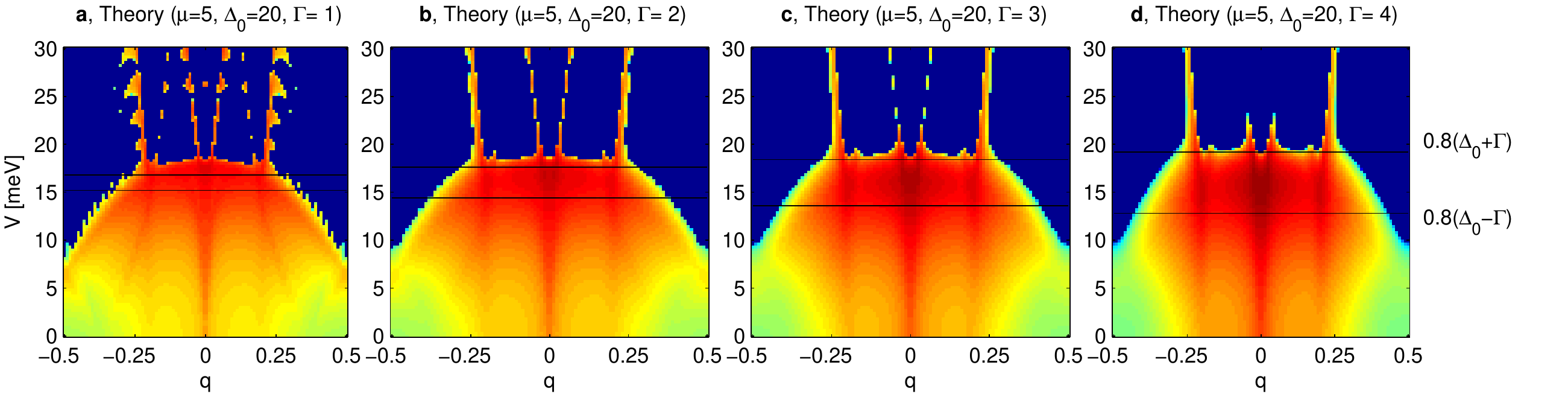}
\caption{{\bf Interplay of dispersive and non-dispersive peaks as function of the inverse quasiparticle lifetime $\bf \Gamma$}. Same as Fig.~\ref{fig:25_comp}{\bf a} for increasing values of the gap and inverse quasiparticle lifetime (corresponding to increasing underdoping). Sharp non-dispersive features are seen at voltages $V>0.8 (\D_0 +\Gamma_0)$ and dispersive features at $V<0.8 (\D_0 -\Gamma_0)$. 
}
\label{fig:2ter_theory}
\end{figure}

\myparagraph{Emergence of two energy scales in STM experiments}
\label{sec:twoscales}
STM measurements of Bi2212 show dispersive features at low voltages and sharp non-dispersive ones at high voltages. As noted in Refs.~ \cite{alldredge08,schmidt11,fujita11}, the transition between these two regimes is interrupted by an intermediate voltage interval in which neither dispersive nor sharp non-dispersive peaks are observed. The size of this region increases with underdoping, giving the impression of two independent energy scales. In Fig.~\ref{fig:2ter_theory} it is shown that the transition between the different regimes corresponds approximately to $0.8(\Delta_0-\Gamma)$ and $0.8(\Delta_0+\Gamma)$. According to this interpretation, the second energy scale observed in STM measurements is related to the quasiparticle lifetime, rather than a distinct energy gap.

\myparagraph{Homogeneous component of the STM signal}
\label{sec:homo}

\begin{figure}[b]
\centering
\begin{tabular}{c c c}
\includegraphics[scale=0.8]{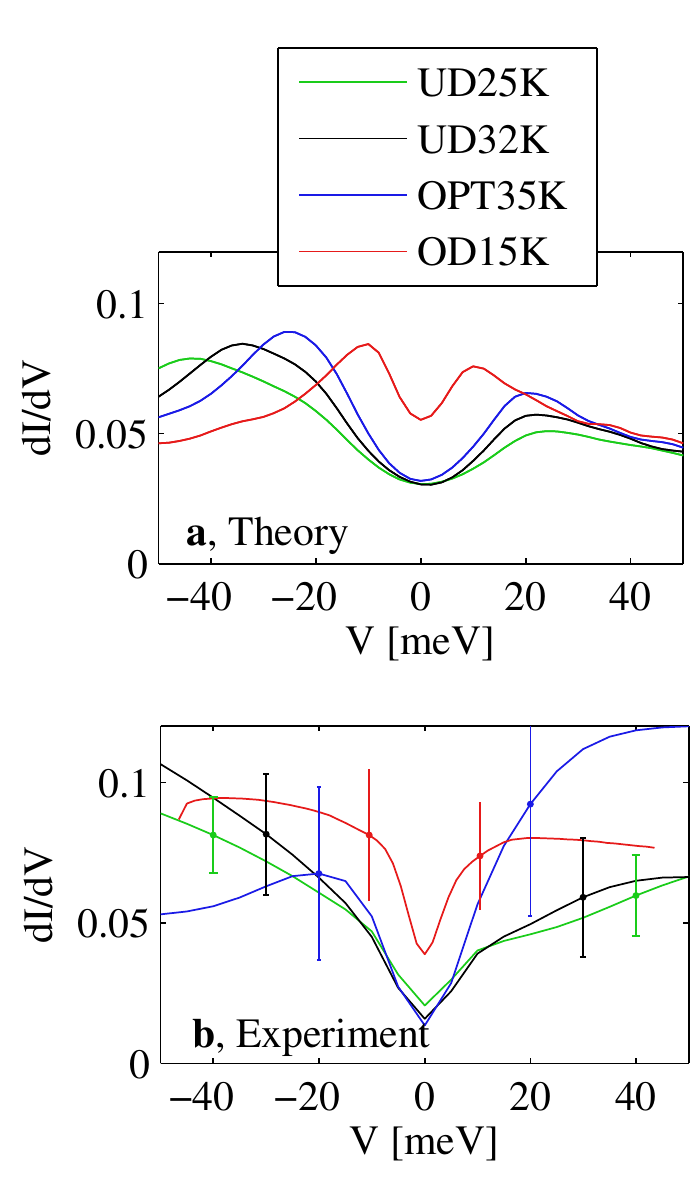} &
\includegraphics[scale=0.82]{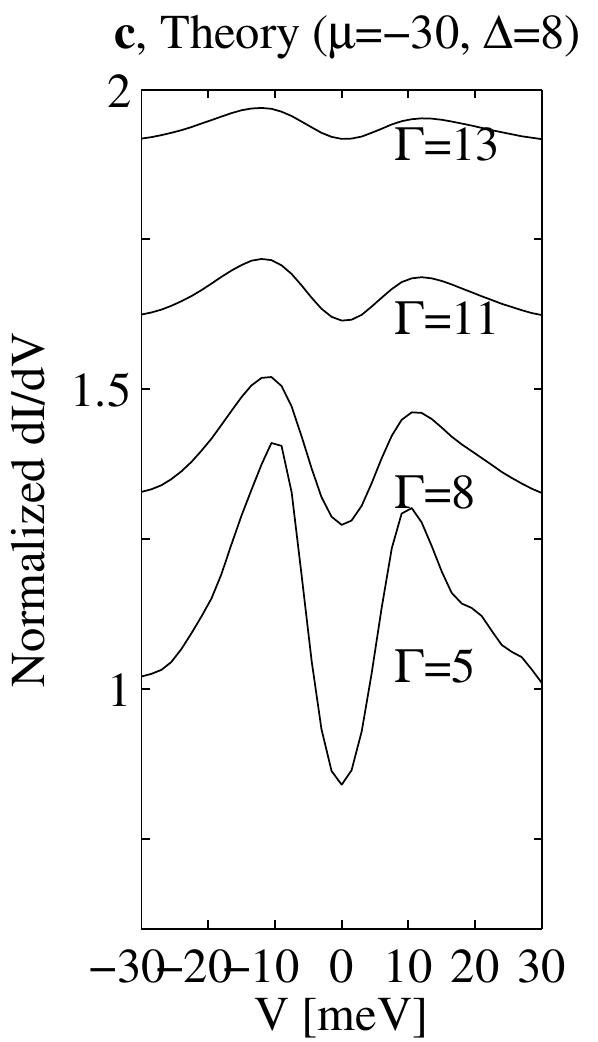} &
\includegraphics[scale=0.5]{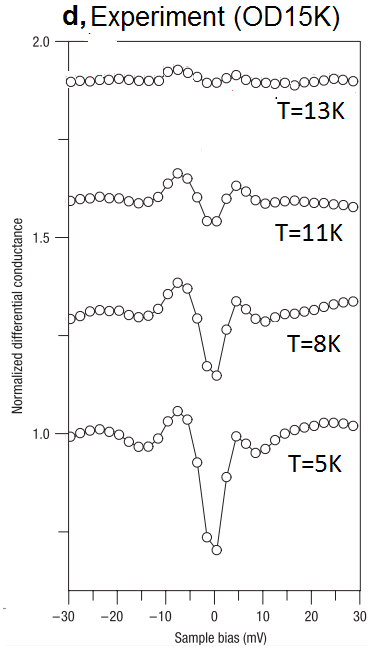} 
\end{tabular}
\caption{{\bf Homogeneous component of the LDOS in Pb-Bi2201. } {\bf a-b}, Theoretical calculation and experimental measurement of the homogeneous component of the differential conductance in four samples of Bi2201. The theoretical calculations were performed using the parameters of Table \ref{table1}. The large error bars in the experimental measurements are due to the inhomogenous component.  {\bf c,} Theoretical predictions of the normalized conductance (see text) for increasing values of $\Gamma$ (all energies are given in meV). {\bf d,} Experimental measurements for an overdoped sample of Pb-Bi2201 (OD15K), at different temperatures. Adapted from Ref.~\cite{hudson07}. Each curve is shifted by 0.3 in the vertical direction for clarity.}\label{fig:0_comp}
\end{figure}

In this section we consider the spatially-homogenous conductance $dI/dV$. In actual experiments, this component can be measured by averaging the STM signal over a large area $\Omega$:
\be\frac{dI}{dV} =\frac1\Omega \int_\Omega dr~\frac{dI(r,V)}{dV} \equiv g(q=0,V)=\sum_k{\rm Im}[G(k,\omega)]\label{eq:homo}\ee 
The theoretical predictions and experimental observation of this component are compared in  Fig.~\ref{fig:0_comp}{\bf a-b} and show a good agreement, within the large error bars of the experimental observations. These error bars are due to the inhomogenous component due to the scattering from local impurities and can be reduced only by averaging over larger areas. At positive voltages, the theoretical curves show a maximum in correspondence of the superconducting gap, $V_+\approx\Delta_0$. At negative voltages the signal shows a broad maximum at the doping-dependent voltage $V_-\approx - 40{\rm meV} -\mu$. This maximum, which is simply due to the particular form of the band structure, could create the impression of gap that increases with underdoping. For any given point in space, the actual STM spectrum is the sum of the homogenous component (\ref{eq:homo}), peaked at $V_+$ and $V_-$ and an inhomogenous contribution (\ref{eq:rhoqw}), whose actual maximum varies in space. The sum of these two terms is expected to give rise to a ``kink'', which was indeed universally observed in experiments (see for example Ref.~\cite{howald01}).

An alternative method to extract the homogeneous component of the STM signal has been proposed in Ref.~\cite{hudson08}. Assuming that in the normal phase $g(x,V)={\rm const}$, the homogenous component of the differential conductance can be obtained from $g(x,V,T)/g(x,V,T_{\rm norm})$, where $T_{\rm norm}>T_c$ is an arbitrary temperature. The experimental signal is reproduced in Fig.~\ref{fig:0_comp}{\bf c} for an overdoped sample of Pb-Bi2201 with $T_c=15$meV, using $T_{\rm norm}=17$meV. The position of the peaks coincides with our identification of the superconducting gap for this sample, $\D_0=8$meV. As the temperature increases, the distance between the peaks does not significantly vary, but their visibility rapidly diminishes. %This effect is in line with previous works on disordered interacting systems  \cite{Lee1985}, predicting that $\Gamma$ should be approximately proportional to the temperature (up to logarithmic corrections). 
In Fig.~\ref{fig:0_comp}{\bf c} we reproduce this result by assuming a linear dependence between $\Gamma$ and the temperature. (In our case, we have established that, at $T=6K$, the inverse quasiparticle lifetime $\Gamma=6$meV, leading to the simple relation $\Gamma/T\approx1$meV/K, see also Appendix \ref{sec:phasediagram}). Using this assumption and normalizing the theoretical predictions with respect to the value at $\Gamma=17$meV, we obtain a good agreement between theory and experiment, as shown in Fig.~\ref{fig:0_comp}{\bf c-d}.

\myparagraph{Analysis of the non-dispersive peak at $\bf q_{\pi,\pi} =(0.5,0.5)\times 2\pi$ }
\label{sec:Qpipi}

\begin{figure}[b]
\centering
\includegraphics[scale=0.8]{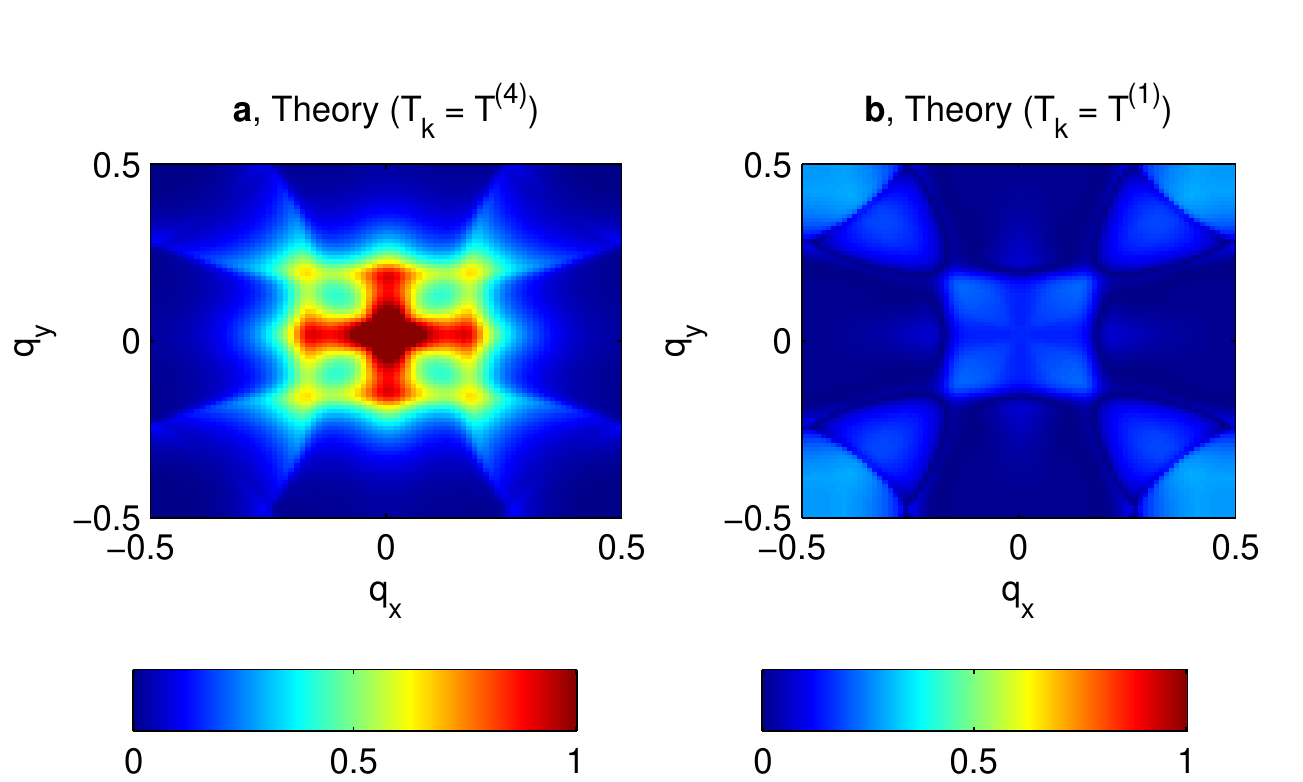}
\includegraphics[scale=0.8]{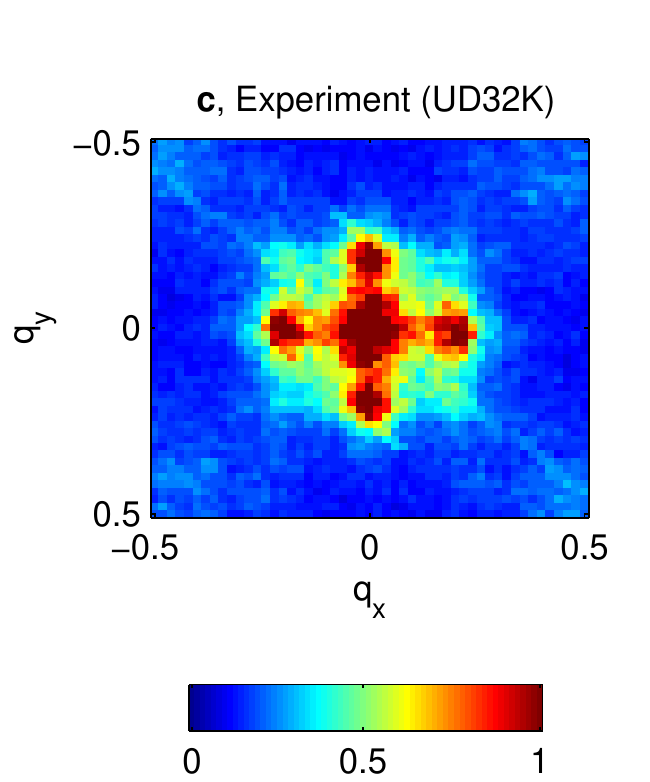}
\caption{{\bf STM signal at $\vec q=(q_x,q_y)\times 2\pi$ and fixed voltage $V=5$meV}. {\bf a}, Theoretical predictions, for local modulations of the gap, Eq.~(\ref{eq:rhoqw}) with $T_k = T^{(4)}_k$. {\bf b}, Same as before, for local modulations of the chemical potential, $T_k=T^{(1)}_k$. {\bf c}, Experimental signal $g(q,V)$ for the an underdoped sample (UD32K). The intensity peak at $q_{\pi,\pi}=(0.5,0.5)\times 2\pi$ observed in the experiment is due to modulations of the chemical potential.}\label{fig:6_comp}%singleband_v7_fig1
\end{figure}

\begin{figure}[p]
\centering
\includegraphics[scale=0.65]{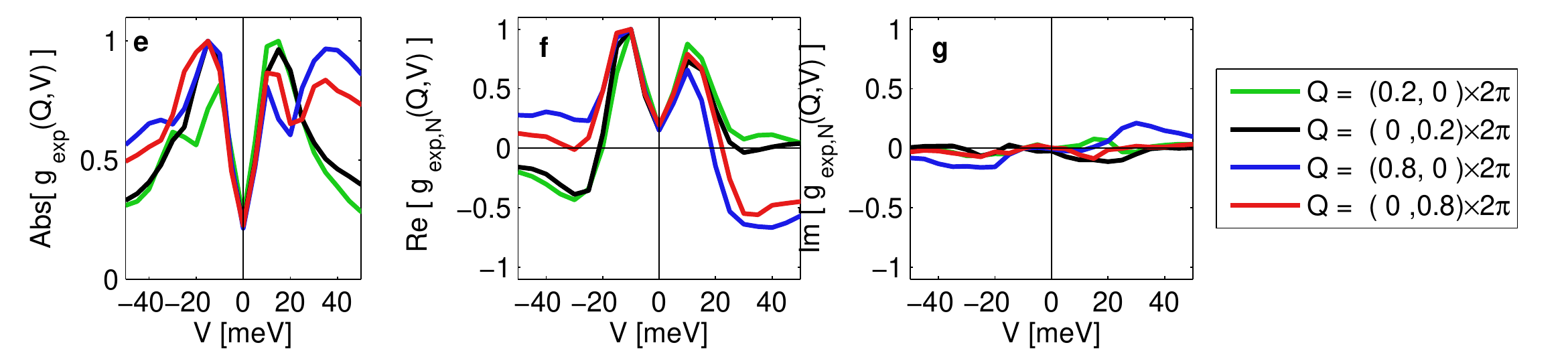}
\includegraphics[scale=0.65]{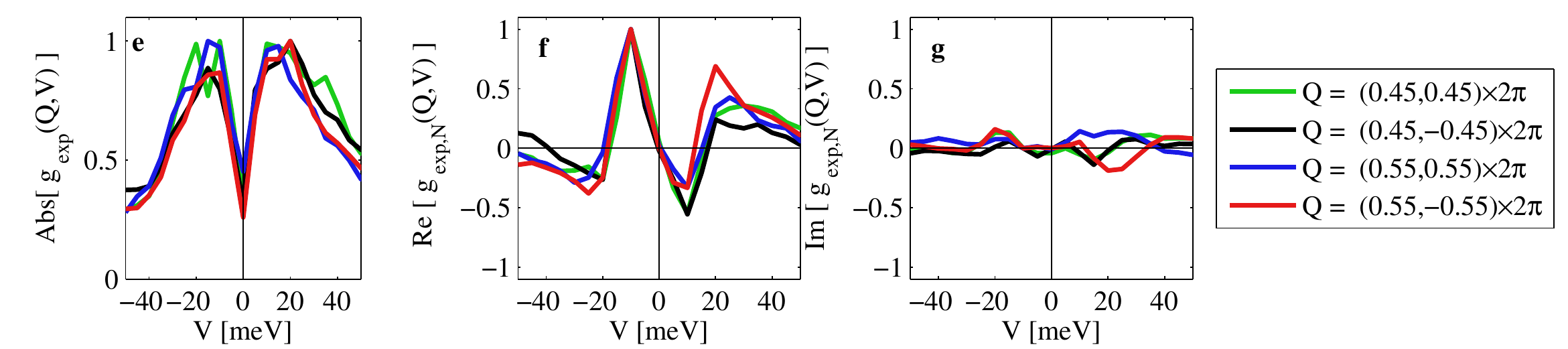}
\caption{{\bf Voltage dependence of the experimental signal at different wavevectors.} Same as Fig.~\ref{fig:1_comp}{\bf a-c} for the optimally doped sample OPT35K and $e^{i\phi(q)}=\rho(q,V$=$-10$meV$)~/~|\rho(q,V$=$-10$meV$)|$. The Fourier component at small wavevectors is predominantly symmetric (gap modulations), while the Fourier component around the $q_{\pi,\pi}$ peak is predominantly antisymmetric (charge modulations).} %This observation is consistent with Fig.~\ref{fig:6_comp}, showing that $g_{\rm th}^{(4)}$ is larger (smaller) than $g_{\rm th}^{(1)}$ at small (large) momenta.}
\label{fig:1bis_experiment}
\end{figure}

Figure \ref{fig:6_comp} shows a two-dimensional cut of the data at fixed voltage $V=5$meV, for wavevectors inside the first Brillouin zone. Comparing subplots {\bf a} and {\bf c} we find that the theory quantitatively reproduces the experiment, with one important exception: the experiment shows a broad peak around $q_{\pi,\pi}\equiv(\pm 0.5,\pm0.5)\times 2\pi$, while the theoretical predictions exactly vanishes there (due to the symmetry of the coherence factors appearing in Eq.~(\ref{eq:rhoqw})). To identify the nature of the $q_{\pi,\pi}$ peak, we study its voltage dependence (Fig.\ref{fig:1bis_experiment}) and find it to be anti-symmetric with r	espect to the $V\to-V$. As explained above, this behavior is characteristic of the scattering from local modulations of the chemical potential.  Fig.~\ref{fig:6_comp}{\bf c} shows that, indeed, this type of perturbation leads to a g-map that is peaked around $q_{\pi,\pi}$. Our findings may also explain the experimental observations of Ref.~\cite{yanghe13}, who showed that the peak at $q_{\pi,\pi}$ responds to magnetic field and temperature in the opposite way than the rest of the map, highlighting its different physical origin.

\myparagraph{Effects of the band structure on the REXS signal}
\label{sec:REXS}

\begin{figure}
\includegraphics{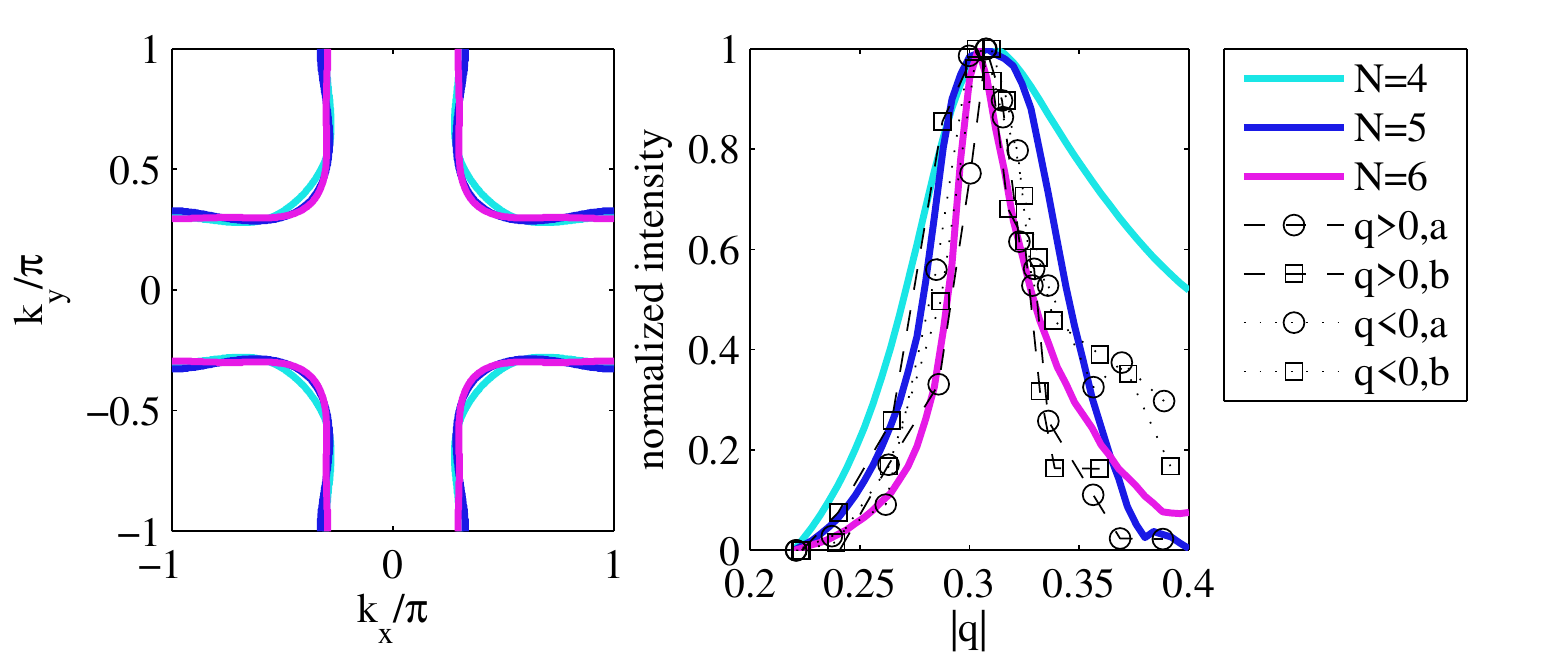}
\caption{{\bf Wavevector dependence of the REXS signal of Y123 at $\vec q=(q,0)\times 2\pi$.} {\bf a}, Fermi surface associated with three different phenomenological models, including $N$ fitting parameter (see Table \ref{table2}). The model $N=6$ is obtained by adding an additional arbitrary term to the band structure of Ref.\cite{shen98} (and adjusting the chemical potential to keep the density fixed). {\bf b}, Predicted and observed REXS signal. The discrepancy between theory (see text) and experiment (\cite{ghiringhelli12}) is far below the combined uncertainty of both. The theoretical curves where obtained using the same parameters as in Fig.~\ref{fig:REXS}. To allow a comparison between the different models and experiments, we have renormalized each curve by subtracting the value at $q=0.22$ and dividing by the maximal intensity. }
\label{fig:REXS2}
\end{figure}

In the main body of the article we found that the predicted width of the REXS peak in Y123 is larger than the one observed in experiments \cite{ghiringhelli12}. 
One interesting possibility is that this discrepancy is due to an enhancement of the CDW order caused by electron-electron interactions. This effect can be described using an random-phase approximation (RPA) \cite{podolsky03,kivelson03} and, in general, acts to sharpen the predicted peak. Here we follow a simpler interpretation and relate the observed discrepancy in the REXS signal to a deviation of the actual band structure from the phenomenological model obtained in Ref.~\cite{shen98}. Unlike the case of Bi2212, the band structure of Y123 is known less accurately, due to surface effects and to the presence of CuO chains \cite{pasani10}. In Fig.~\ref{fig:REXS2} we compare calculations for the REXS signal using two different phenomenological band structures with similar Fermi surfaces. The position of the REXS peak $q=0.31$ is uniquely determined by the doping, and is largely model independent. In contrast, the widths of the predicted signals significantly differ between the two models and vary from $\xi=0.1$ to $\xi=0.07$. We note that the phenomenological model with a larger number of parameters ($N=5$) displays a sharper peak and offers a  better agreement with experiments. In general, a sharp peak in the REXS signal requires a nested band-structure, whose characterization involves many fitting parameters. Accordingly, we observe that the band structure of Ref.~\cite{yagi10}, obtained using one single fitting parameter, does not generate any significant REXS peak. To further explore this point we consider the effects of an additional momentum-dependent term in the band structure of Ref.~\cite{shen98}, with approximately the same amplitude as the previous ones (see last column of Table \ref{table2}). We find that this term leads to a further sharpening of the REXS peak and an excellent agreement with experiments. We therefore propose that REXS experiments can be used to probe the band structure of the antinodal regions of cuprates.

\begin{table}[p]
\begin{tabular}{|c | r r r r r |}
\hline
 Material & Bi2212 & Y123(A) & Y123(B)  & Y123(B) & Y123(B)\\
Number of fitting paramaters & N=5 & N=5 & N=5 & N=4 & N=6 \\
Reference &  \cite{campuzano95}&  \cite{shen98} & \cite{shen98} & \cite{pasani10} & \\
\hline
1
  &    0.1305  &    0.4368  &    0.1756  &    0.0500  &    0.1456
\\$\frac12(\cos k_x +\cos k_y)$ 
  &   -0.5951  &   -1.0939  &   -1.1259  &   -0.4200  &   -1.1259
\\$\cos k_x \cos k_y$ 
  &    0.1636  &    0.5612  &    0.5540  &    0.1163  &    0.5540
\\$\frac12(\cos 2 k_x + \cos 2 k_y)$ 
  &   -0.0519  &   -0.0776  &   -0.1774  &   -0.0983  &   -0.1774
\\$\frac12(\cos 2k_x \cos k_y + \cos k_x \cos 2 k_y)$
  &   -0.1117  &   -0.1041  &   -0.0701  &   -0.0353  &   -0.0701
\\$\cos 2k_x \cos 2k_y$ 
  &    0.0510  &    0.0674  &    0.1286  &         0  &    0.1286
\\$\cos 2k_x \cos k_x + \cos 2k_y \cos k_y$ 
  &         0  &         0  &         0  &         0  &    - 0.1000
\\\hline
\end{tabular}
\caption{Phenomenological band structure of Bi2212 (used for calculations of Pb-Bi2201 as well) and Y123 (bonding (B) and antibonding (A)). The first four columns were obtained from least square fits of ARPES measurements and used without modifications in the present analysis. Ref.~\cite{pasani10} further analyses the effects of intraplane couplings, which are neglected in the present analysis. The last column is obtained from the model of Ref.~\cite{shen98}, by adding an additional momentum-dependent term to the band structure and correcting the chemical potential to conserve the area of the Fermi surface.}\label{table2}
\end{table}

\begin{figure}
\includegraphics[scale=0.8]{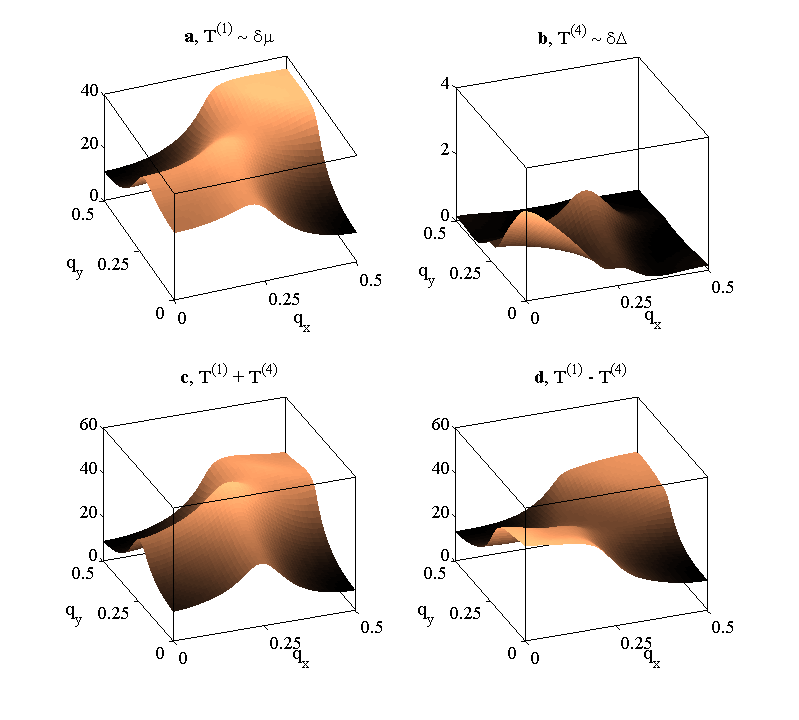}
\caption{{\bf Predicted REXS signal as a function of the two-dimensional wavevector} $\mathbf{\vec q = (q_x, q_y)\times 2\pi}$. The physical parameters refer to an underdoped sample of Bi2212 (see the inset of Fig.~\ref{fig:REXS}). Each subplot describes the effects of a different type of local modulation (see text).}
\label{fig:REXS2Dbis}
\end{figure}

\myparagraph{REXS signal in two dimensions}
In Fig.~\ref{fig:REXS2Dbis} we plot the predicted REXS signal as a function of the two-dimensional wave-vector $\vec q$. Each subplot corresponds to a different type of local scatterer: {\bf a}, local modulations of the chemical potential; {\bf b}, local modulations of the pairing gap; {\bf c}, the sum of the previous two; and {\bf d}, their difference. See Methods section for the definition of the corresponding scattering matrices $T^{(\a)}$. By comparing subplots {\bf a} and {\bf b} we observe that REXS experiments couple more strongly to the modulations of the chemical potential, than to modulations of the pairing gap. This effect is due to the integration over frequencies appearing in Eq.~\ref{eq:REXS}: as shown in Fig.~\ref{fig:1_comp}, the differential conductance induced by a modulation of the pairing gap ($T^{(4)}$) is very small for any $\omega>\Delta_0$, while the effects of modulation of the chemical potential survives far above $\Delta_0$. Exploiting the results obtained from the analysis of the STM signal (Appendix \ref{sec:Qpipi}), we conjecture that subplot {\bf c} should best reproduce the physical situation. In addition to the peaks at $\vec q_a=(0.25,0)\times 2\pi$ and $q_b=(0,0.25)\times 2\pi$, we predict a pronounced peak at $\vec q_a \pm \vec q_b$, whose maximal intensity is larger than the one predicted for $\vec q_a$ and $\vec q_b$.

\myparagraph{Implications for the phase diagram}
\label{sec:phasediagram}
Our analysis indicates that a finite quasiparticle lifetime is fundamental for understanding the single-particle properties of cuprates. Here we explore the possibility that $\Gamma$ may also play an important role in determining the critical temperature $T_c$. We observe that, in Pb-Bi2201, $T_c$ seems to correspond to the point were antinodal quasiparticles become over-damped, i.e. where their inverse lifetime equals to twice their gap:  $\Gamma(T_c)=2\Delta_0$. To obtain this result, we assume a linear dependence of $\Gamma$ on the temperature, $\Gamma(T) = \alpha T$, found in both theoretical calculations  \cite{dahm95,pao95,ossadnik08} and experiments  \cite{hussey06}.  Starting from the observed values of $\Delta_0$ and $\Gamma$ (see Table \ref{table1}, obtained from STM measurements at $T=6K$) and requiring $\alpha T_c=2\Delta_0$, we obtain $T_c = (26\pm2)K,(30\pm3)K,(31\pm3)K,(16\pm2)K$, consistent with the actual values  $T_c=25K,32K,35K,15K$. In optimally-doped Bi2212 the gap is 1.5 times larger ($\Delta_0\approx30$meV) and the quasiparticle lifetime 2 times smaller (as can be inferred from the measured value $\Gamma\approx1$meV at $1.9K$  \cite{alldredge08}), leading to a critical temperature that is approximately 3 times larger, $T_c\approx90K$. This phenomenological observation suggests that  the critical temperature could be further increased by decreasing $\Gamma$. The opposite effect (i.e. a decrease of $T_c$ for increasing $\Gamma$) has been recently demonstrated in experiments~ \cite{dessau13}.

\begin{figure}[p]
\centering
\begin{tabular}{c}
\includegraphics[scale=1]{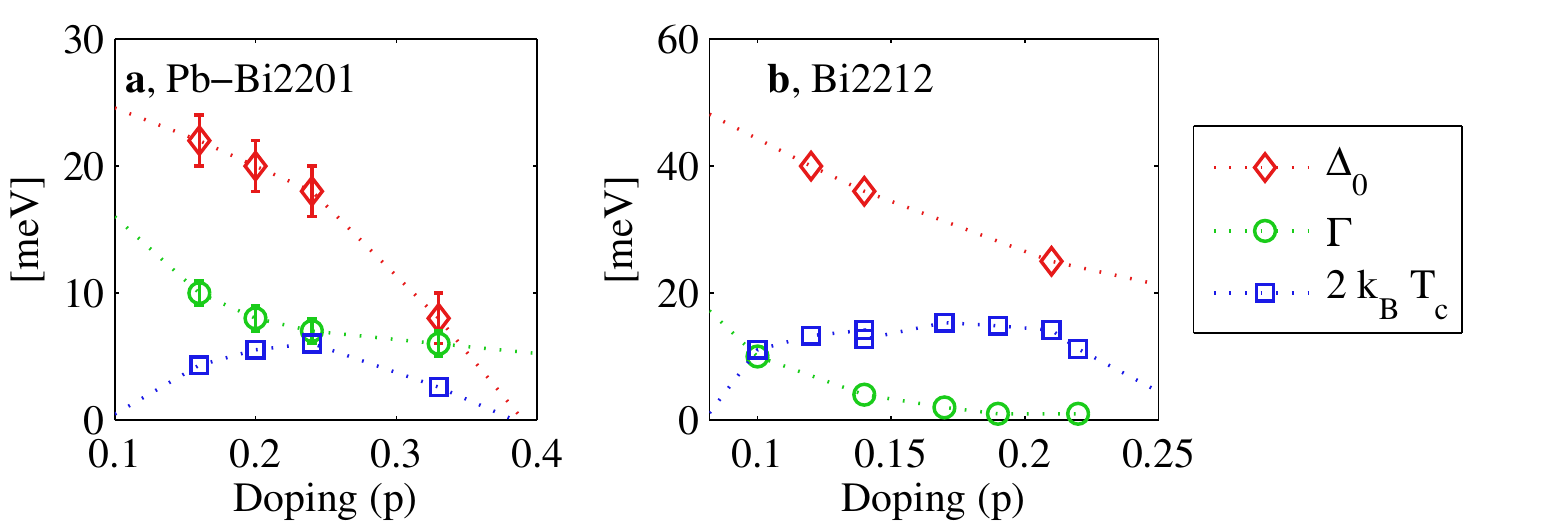}\\
\includegraphics[scale=0.8]{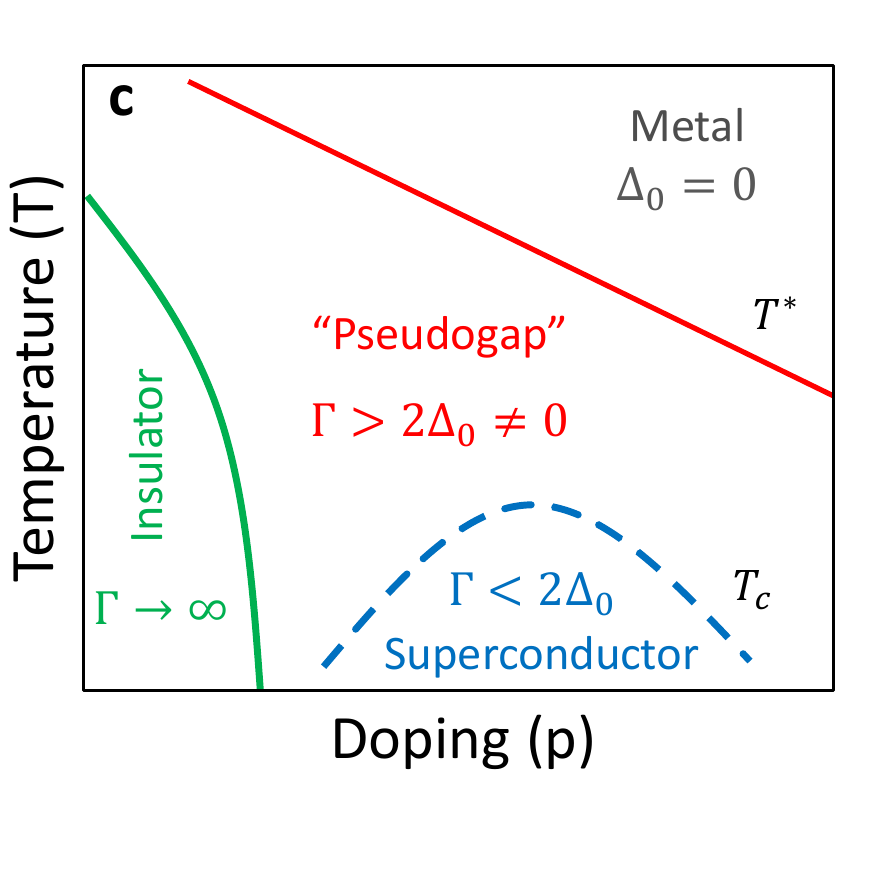}
\end{tabular}
\caption{{\bf Evolution of the model parameters as function of doping and temperature}.  %The transition between the metallic phase and the ``pseudogap'' phase corresponds to the onset of the superconducting gap $\Delta_0$. The latter phase is distinguished from the superconducting phase only due to a larger inverse quasiparticle lifetime $\Gamma>2\Delta_0$. 
{\bf a}, Evolution of the superconducting gap $\D_0$, inverse quasiparticle lifetime $\Gamma$, and critical temperature $T_c$ in 4 samples of Pb-Bi2201 (Table \ref{table1}. {\bf b}, Same as {\bf a} for 8 samples of Bi2212 analyzed by Dipasupil \etal  \cite{dipasupil02} and Alldredge \etal  \cite{alldredge08}. {\bf c}, Proposed phase diagram of Pb-Bi2201. %A finite $d$-wave gap in the pseudogap regime is also ``clearly'' observed in ARPES measurements ``as the location of the leading-edge midpoint'' (quoting Ref.~\cite{ARPES_review}, page 518)
}\label{fig:doping}
\vspace{-0.5cm}
\end{figure}

\begin{figure}[p]
\includegraphics[scale=0.7]{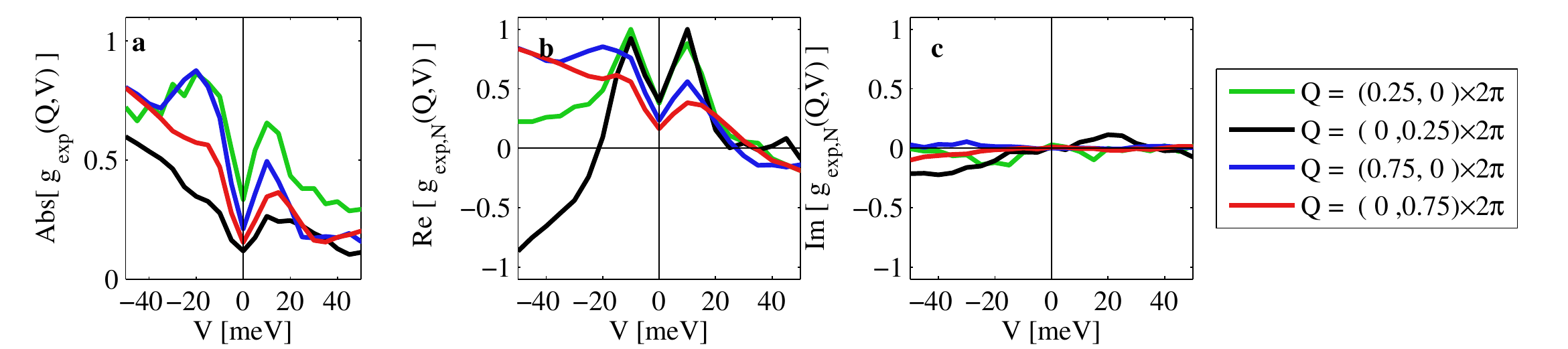}
\includegraphics[scale=0.7]{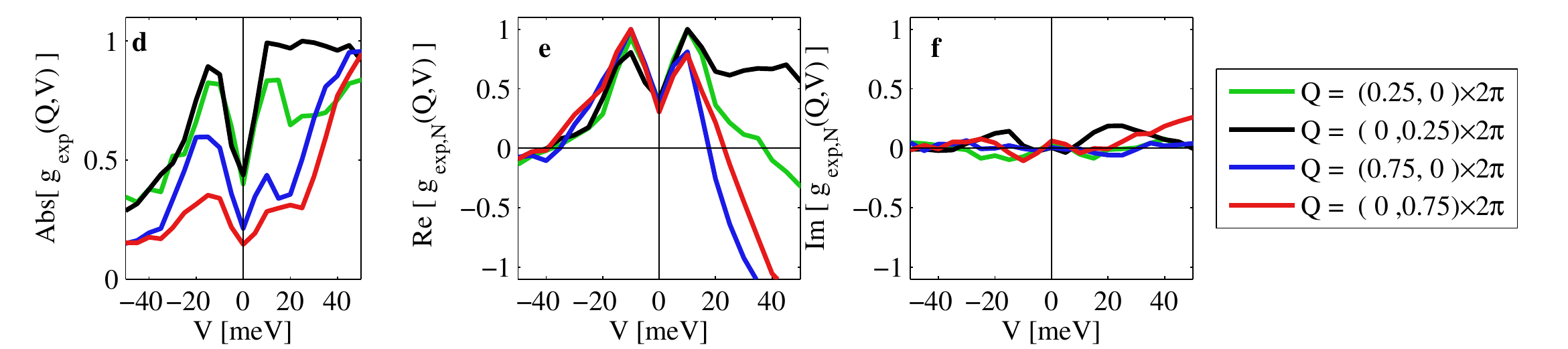}
\caption{{\bf Effects of the different normalizations on the experimental signal}. Same as Fig.~\ref{fig:1_comp}{\bf e-g} but with a different normalization (see text). {\bf a-c}, The spectrum at each point in real space is divided by $I_{\rm max}(r)$. {\bf d-f}, Each point is divided by $I_{\rm min}(r)$.}\label{fig:1bis_exp}
\vspace{-0.5cm}
\end{figure}

\myparagraph{Normalization of the STM data}
\label{sec:normalization}
One main technical difficulty in performing STM experiments is related to the unkown distance between the tip and the sample, which can vary from point to point. To overcome this problem, the experimental data is usually normalized at each point by the current at the maximal observed voltage $I_{\rm max}(r)=I(r,V_{\rm max})=\Delta V \sum_{0}^{V_{\rm max}} dI(r,V)/dV$. Alternative normalization procedures include dividing by the current at the minimal voltage $I_{\rm min}(r)=I(r,V_{\rm min})=\Delta V \sum_{V_{\rm min}}^0 dI(r,V)/dV$, or by the difference $I_{\rm max}(r)-I_{\rm min}(r)$. This third normalization was used in generating the plots of Fig.~\ref{fig:1_comp}{\bf e-g} because it does not introduce spurious asymmetries between positive and negative voltages. The same plots, but with different normalizations are shown Fig.~\ref{fig:1bis_exp}. In subplots {\bf a} and {\bf d} we observe that the normalization procedure does not significantly affect the absolute value of the signal at small wavevectors (black and green curves), but radically changes the signal at large wavevectors (red and blue curves). These changes are mitigated by splitting the signal into its real and imaginary components (subplot {\bf b-c}, {\bf e-f}). In particular, we observe that the peak at $V\approx 50$meV, observed in the absolute value of the large-wavector signal  (curves red and blue of subplot {\bf d}) is actually a local minimum, associated with a change in sign of the real signal at $V\approx\Delta_0=20$meV.

%\bibliographystyle{unsrt}
%\bibliography{STM}

\end{document}